\documentclass[epj,color,sorting=none]{svjour}
\usepackage{graphics}
\usepackage{float}
\usepackage{color}
\usepackage{bm}% bold math
\usepackage{amsmath,amssymb}
\usepackage[normalem]{ulem}

\DeclareMathAlphabet\mathbfcal{OMS}{cmsy}{b}{n}
\begin{document}
\title{Active dumbbells: dynamics and morphology in the coexisting region}
%\subtitle{Dynamics and morphology  of the coexisting phase of an active  molecular system}
\author{Isabella Petrelli\inst{1}, Pasquale Digregorio\inst{1}, Leticia F. Cugliandolo\inst{2}, Giuseppe Gonnella\inst{1}, Antonio Suma\inst{3}% etc
}                     % Do not remove
%
%\offprints{
%isabella.petrelli1@gmail.com}          % Insert a name or remove this line
%
\institute{Dipartimento di Fisica, Universit\`a degli Studi di Bari and INFN,
Sezione di Bari, via Amendola 173, Bari, I-70126, Italy
\and 
Sorbonne Universit\'e, Laboratoire de Physique Th\'eorique et Hautes Energies,
CNRS UMR 7589, 4 Place Jussieu, 75252 Paris Cedex 05, France
\and 
SISSA - Scuola Internazionale Superiore di Studi Avanzati, Via Bonomea 265, 34136 Trieste, Italy
and  Institute for Computational Molecular Science, College of Science and Technology, Temple University, Philadelphia, PA 19122, USA}
\date{Received: date / Revised version: date}
\abstract{
With the help of molecular dynamics simulations we 
study an ensemble of active dumbbells
in purely repulsive interaction. We derive the phase diagram in the density-activity plane and we
characterise the various phases with liquid, hexatic and solid character. The analysis of the structural and dynamical properties, 
such as enstrophy, mean square displacement, polarisation, and correlation functions, shows the continuous character of liquid and 
hexatic phases in the coexisting region when the activity is increased starting from the passive limit. 
 % end of PACS codes
} %end of abstract
\authorrunning{Petrelli {\it et al.}}
\titlerunning{Active dumbbells}
\maketitle
\section{Introduction}
\label{intro}

We open this article with a description of a number of noticeable facts of active, but also passive, matter that
have motivated our studies of an ensemble of self-propelled dumbbells in purely repulsive interaction
confined to  move in a two dimensional space.
 
\subsection{Non-equilibrium dynamics under local bulk energy injection}

{\it Active materials} are many-body systems
composed of self-driven units that convert stored or ambient free energy into systematic movement.
%These model systems capture the essential features of the dynamics of living systems on many scales, ranging from 
%microorganisms to birds or fish.
They are, typically, living systems, and the size of their elements range over many scales, from microorganisms or cells to birds or fish.
Artificial realisations, sometimes easier to control in experiments, have also been designed and  include 
self-propelled colloids~\cite{Palacci10,Theurkauff2012,Bricard13,Lowen}, nanorods~\cite{Paxton04}, droplets~\cite{Thutupalli11,Sanchez12} and active gels made by cytoskeleton extracts  in presence of molecular motors~\cite{Surrey01,Bendix08}
as well as vibrated mechanical walkers~\cite{Narayan07,Kudrolli08,Deseigne10,Deseigne13}. 

From the point of view of physics, active materials are novel and very interesting objects of study. They are 
maintained out of equilibrium by the continuous injection of energy at a microscopic
scale  within the samples. The microscopic dynamics explicitly break detailed balance 
and, in consequence, no equilibrium theorem can be used as a guideline to understand the 
macroscopic behaviour. Nevertheless, since the consumed energy is partly dissipated 
into the medium, different non-equilibrium steady states establish and they are still amenable to  be 
studied with physics tools. Phase transitions between such steady states are possible even in low dimensional cases. The full 
characterisation of the dynamic phase diagram and the various phases is one of the issues that attracts physicists' attention.

\subsection{Density, form, and dimensionality}

The dynamics of systems as seemingly unrelated as flocks of birds, swarms of bacteria and vibrated rods share many features in 
common and, up to a certain extent, they can be treated within 
the common theoretical framework of active fluids or suspensions~\cite{Fletcher09,Menon10,Ramaswamy10,Vicsek12,Marchetti13,Cates12,Romanczuk12,Safran13,Marenduzzo14,Elgeti15,gonn15}.
Different approaches aimed at a coarse-grained description based on general symmetry arguments are available but 
fluctuations and phase transitions have been especially analysed in the context of agent-based models.
Although many papers study motion in the dilute limit, much less is known about the behaviour of {\it dense} ensembles 
subject to  not so strong activity. In particular, the connection with the passive limit, and their own complex phases and 
phase transitions, have not been studied in so much detail.

Natural and artificial active matter are rarely constituted by spherically symmetric elements. On the contrary, 
{\it anisotropic} objects are much more common.  Examples of living realisations in which the constituents 
have anisotropic shapes include reconstituted layers of 
confluent epithelial cells~\cite{Park15} and biofilms formed by dense collections of rod-shaped 
bacteria~\cite{Lambert14}. Artificial cases also exist and an example is given by shaken non-spherical grains~\cite{Kumar14}. 

For concreteness,
let us focus on the example of a bacterial colony, a many body system made of elongated constituents.
Bacterial colonies occupy three dimensional spaces but they  can also be confined to {\it two
dimensions} when deposited on agar plates.
The collective motion of such systems was studied by biologists since long ago with the aim of understanding 
the chemotaxis and sensory transduction mechanisms~\cite{Koshland1981}. More recently, physicists
have entered this field and, in particular, they have identified 
dynamic phase transitions, order parameters, 
topological defects, {\it etc.}~\cite{Mendelson99,Dombrowski04,Cisneros2007,Sokolov07,Zhang09,Dullens}.
Similar experimental studies were carried out in 
cell populations~\cite{volfson,Riedel05,szabo2006}. 

The way in which bacteria swim, via a dipolar or multipolar field,  is very different from the one in which macroscopic animals do.
At micron length scales, inertial effects are negligible compared to viscous forces. Effectively, the 
swimmers evolve in a {\it zero Reynolds number} limit.     
We can  therefore ensure that dense systems of an\-iso\-trop\-ic self-propelled elements confined to two dimensions,
in the low Reynolds number limit,
are apt to mimic swarms of bacteria and present interesting challenges from the material science point of view.
Before continuing with the description of some general features of active and passive materials, 
we announce that the subject of this paper is the study of  
a minimal model for such systems, where the dynamics of the solvent is neglected and only  interactions with a thermal substrate are
considered. 
Similar models have been used, for instance, in Refs.~\cite{Wensink2012a,Wensink2012b,Hinz2015,Tung2016,Siebert2017,valadares2010catalytic,thakur2012collective,ruckner2007chemically,colberg2017many}
and by our group~\cite{Suma13,Suma14,Suma14b,Suma14c,Suma15a,Suma16,Cugliandolo2017}. We will 
refer to the results of~\cite{Suma13,Suma14,Suma14b,Suma14c,Suma15a,Suma16,Cugliandolo2017} and also present new findings, 
the interpretation of which we highlight in itemised form, in this article.

 \subsection{Motility induced phase separation (MIPS)}

One of the salient features of active matter is its ability to cluster in the absence of cohesive forces.
 
Clustering in Vicsek type models~\cite{vicsek1995novel,czirok1999spontaneously},
induced by explicit orientation  interactions,  and in self-propelled 
hard rods, favoured by the apolar alignment of the anisotropic particles~\cite{Peruani2006,Ginelli2010,Yang2010}, 
was signalled in the early active matter literature.

 Cates {\it et al.}~\cite{Tailleur08,Cates2010,Cates12} argued that clustering and phase separation are generic properties of systems 
 driven out of equilibrium by a persistent local energy input that breaks detailed balance. 
As a consequence, {\it motility induced phase separation} is generated in active materials 
 in the absence of any attractive interaction or breaking of the orientational symmetry.
 
The picture that emerged is  based on an analogy with the liquid-gas transition. The homogeneous gaseous
phase in the active system would become metastable on a binodal line in the (Pe, $\phi$) phase diagram 
and next unstable on a spinodal line. 
(Pe and $\phi$ are dimensionless numbers to be defined later in our model; they represent 
 the strength of the  active force compared with  temperature and particle 
density, respectively.)  Nucleation of the stable (liquid) phase in between these two lines would be 
inhibited by the huge time scales needed, in most cases.
The existence of a critical point at a lowest non-vanishing Pe with liquid-gas coexistence 
was conjectured~\cite{Cates12,Tailleur08,Cates2010,Solon16,Solon18}
({\it e.g.}, found from a linear stability analysis of the homogeneous fluid replacing all interactions by an effective 
propulsion that decreases with density). However, to date, there is no clear-cut evidence of it in numerical studies.

More recently, phase separation was demonstrated in models of soft 
mono-disperse~\cite{Fily12} and poly-disperse~\cite{Fily14}
non-aligning self-propelled disks interacting through harmonic
repulsion (with a cut-off) in  two dimensions. In these papers the onset of freezing (in the polydisperse case) 
and phase separation were determined from the analysis of 
the translational mean-square displacement and the number fluctuations, respectively. 
A bit later,  Redner, Hagan and Baskaran~\cite{Redner13} simulated active Brownian isotropic particles with excluded volume interactions and no alignment. They confirmed phase separation within a coexistence boundary that is similar to the binodal curve of an equilibrium 
fluid with Pe playing the role of the attraction strength. From the analysis of the hexatic order within the clusters at high Pe, these authors 
concluded that the phase separation is between a fluid and an active solid, similar to what is seen near the solid-hexatic transition 
point in a passive two dimensional system (see the next Section). They pointed towards the existence of a critical point at finite Pe, 
and identified nucleation events and coarsening phenomena depending on the depth of the quenches performed.
Finally, we mention  that phase separation was also found in a simple model of Brownian disks in which self-propulsion 
velocities are defined in terms of a persistent Gaussian noise rather than  having  fixed 
norms~\cite{farage15,koumakis14,bettolo15,szamel15,fodor16}.

\subsection{Two dimensional melting}
\label{subsec:2d-melting}

The properties of two dimensional atomic and molecular passive systems and, especially, their melting transition, 
is a fascinating subject that has attracted the attention of  physicists since long ago. Experimental and numerical 
studies have not yet established, beyond any doubt, which are the mechanisms that lead to melting. Two dimensions is 
such a special case since the Mermin-Wagner theorem inhibits the spontaneous symmetry breaking 
of the translational symmetry and, accordingly, positional 
order can only be quasi long-range in bi-dimensional systems. A different type of order, orientational order between the imaginary bonds 
that link the centres of neighbouring particles, can instead be truly long-range in a $2d$ solid.
 
The pioneering numerical study carried out by Alder and Wainwright~\cite{Alder1962} of the $2d$ melting  of a hard disk solid lead to the conclusion that the melting transition is first order even in $2d$. However, around 15 years later, Halperin and Nelson (HN)~\cite{Nelson1979}, and independently Young (Y)~\cite{Young1979}, worked out a theory for $2d$ melting mediated by the unbinding of topological defects. This theory is based on the previous ideas of Kosterlitz and Thouless~\cite{Kosterlitz1973}  (also Berezinskii~\cite{Berezinskii1971a}) and predicts two infinite dimensional phase transitions leading from a solid (with quasi long-range positional and long-range orientational order) to a hexatic phase (with short range positional and quasi long-range orientational order) to the liquid (where both orders are only short-ranged).
The two phase transitions are proposed to be driven by the unbinding of dislocations (solid-hexatic) and the unbinding of disclinations (hexatic-liquid).
Needless to say, the experimental and numerical determination of the transition are difficult and this fact makes the distinction between the two pictures a very interesting issue. The literature on this problem is vast; nevertheless, its full resolution has remained elusive.
   
%%%%   \textcolor{red}{*** L: I propose the following shortening ***}
The two {\it scenarii} should, in principle, be distinguished by the finite size behaviour of the hexatic order parameter in the intermediate phase.
However, its careful determination is quite impossible to make.
The importance of looking at higher moments and correlation functions of this order parameter was
stressed in~\cite{Weber1995a} and the analysis in this paper points towards a first order phase transition
in a hard disk system.

Around 10 years ago, W. Krauth and collaborators came back to this 
problem~\cite{BeKrWi09,BeKr11,Engel13,KaKr15}. In a series of papers, 
they proposed that the transition mechanism depends on the range of the potential,
characterised by its short distance behaviour, say, $U(r) \propto r^{-2n}$.
For relatively smooth potentials with $n\leq 3$ the HNY scenario 
was reproduced by  their numerical simulations. Instead,
for sufficiently hard repulsive potentials ($n>3$) 
the transition was found to be of mixed kind, with the solid-hexatic being as predicted by HNY and the hexatic-liquid
being first order. These numerical studies were performed with an event driven Monte Carlo algorithm that allows one to 
reach very long times and equilibrate systems with unprecedented large sizes. The effect of the interaction potential between 
the elementary constituents
was examined in~\cite{KaKr15}. Spherically symmetric particles were always considered in these papers.

Matter is not only made of symmetric objects. Mol\-e\-cules have non-trivial shapes
and the form of the elementary constituents have non trivial 
effects on the phase diagram and dynamic 
behaviour. The crudest model for a diatomic molecule is a dumbbell made by identical 
beads attached to each other by an edge linking 
their centres. This is usually considered to be  the simplest non-convex body.
The liquid theory of ensembles of such molecules and their numerical investigation 
was the subject of intensive studies, some of them published in 
Refs.~\cite{Talbot1985,Boublik1988,Wojciechowski1992,Goulding1992}.
Certain experimental systems, such as monolayers of diatomic molecules absorbed on crystalline surfaces 
correspond, in an idealised representation, to a bidimensional dumbbell system~\cite{Diehl1980}.
One can easily imagine that dumbbell systems can jam more or less easily depending on 
the relation between the bond length and bead diameter, and the rigidity of the link, and these parameters have an effect on the melting/solidification mechanisms. 

\subsection{MIPS \& Melting?}

\begin{figure}[h!]
\resizebox{\columnwidth}{!}{
  \includegraphics{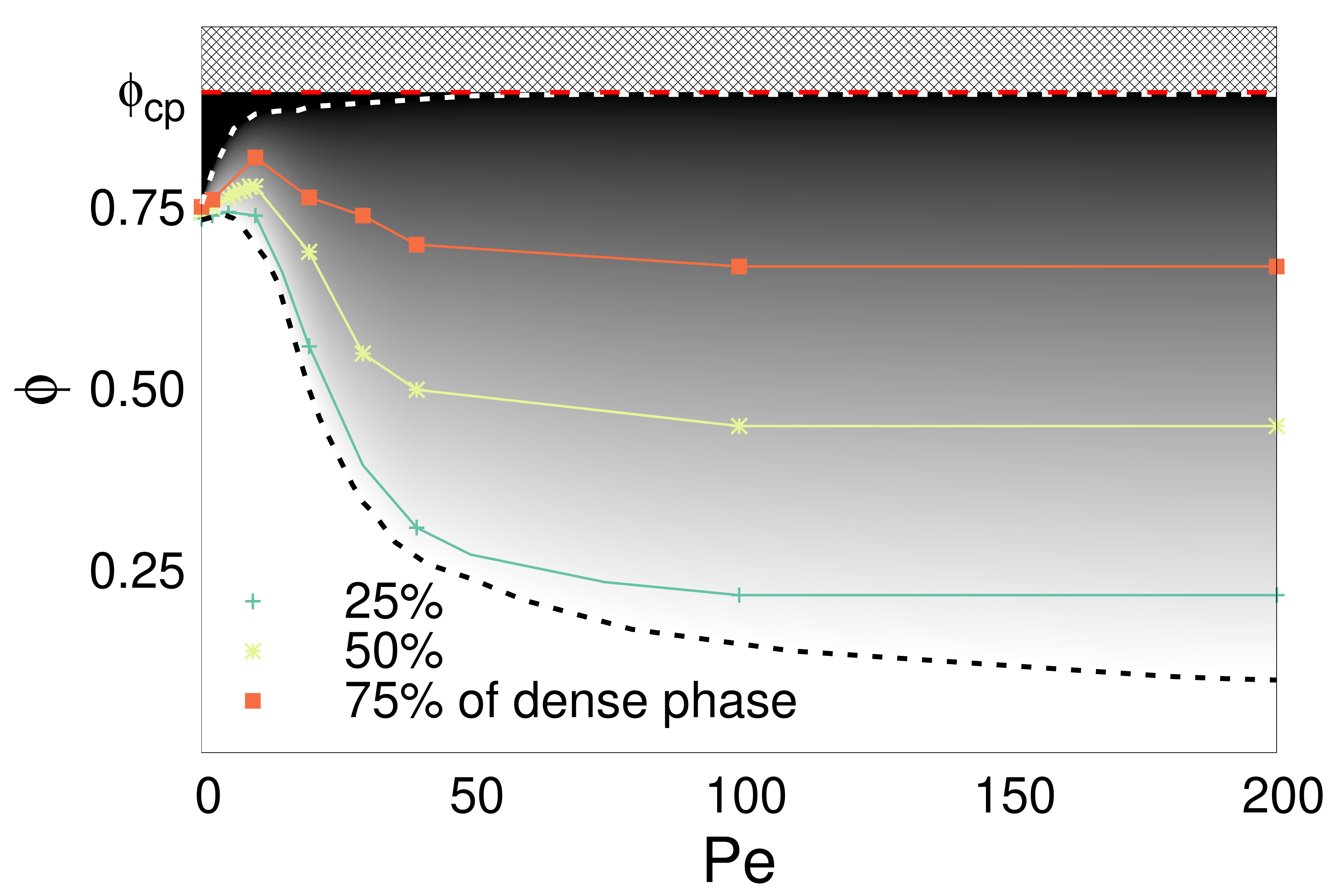}
}
\caption{Phase diagram of the active dumbbell system. The grey scale represents the amount of dense
 ordered phase in the system.
The dotted curves are the location of the lower (in black) and upper (in white) limits of coexistence. In the black region above the white dotted line the system is in a single ordered phase. The solid coloured line-points indicate 
curves on which there is a constant proportion of areas covered by the dense and dilute phases in the 
coexistence region (see the key). The horizontal red dotted line at $\phi_{\rm cp}=0.91$ refers the close packing limit of hard disks.
}
\label{fig:phase_diagram}
\end{figure}

A natural question to ask is whether MIPS and the coexistence of dense and loose regions, with and without 
hexatic order, predicted by the two step melting scenario of Krauth {\it et al.} are continuously connected
under activity or not at all, that is to say, whether they are independent phenomena. This question was asked in~\cite{Cugliandolo2017} for the active dumbbell system and evidence for there being a continuous relation between the two was given in this paper. The resulting phase diagram is announced in Fig.~\ref{fig:phase_diagram} and it will be further discussed in the text. The comparison to recent studies of active disks~\cite{Bialke12,KlKaKr18,DigrLeCuGoPaSu18} in which similar questions are posed will be given in the concluding Section.

This paper contains an extended review of the studies that some of us presented in Refs.~\cite{Suma13,Suma14,Suma14b,Suma14c,Suma15a,Suma16,Cugliandolo2017} 
and new results based on further simulations performed for this work, that complement our analysis of the bidimensional active dumbbell system in 
repulsive interaction. In Ref.~\cite{Cugliandolo2017} the phase diagram of the active dumbbells fluid was shown only for 
Pe ${\leq 40}$, while in this paper
we complement the study exploring a wider region.
We also  present a new detailed analysis of  
structure properties using the orientational and positional correlation functions, and of dynamic properties by analysing the angular velocity of 
clusters and the enstrophy distribution functions, that we correlate to the dumbbells orientation and the polarisation field within these structures.
We confirm that we do not see any discontinuity in the behaviour of structural and dynamic observables within the coexistence region in the phase diagram.

The structure of the paper is the following. In Sec.~\ref{sec:model} we recall the definition of the 
model and we give values of the parameters that we use to obtain the original results presented in this paper.
Section~\ref{sec:single-dumbbell} presents a summary of the behaviour of the single active
dumbbell. 
The two next Sections are devoted to the presentation of our results. In Sec.~\ref{sec:structure}
we focus on the structural properties of the system in its various steady states:
in Subsec.~\ref{sec:observables} we define a handful of 
instantaneous observables and in
Subsec.~\ref{sec:numerics} we show and discuss their numerical measurements.
In Sec.~\ref{sec:dynamic} we define and evaluate several dynamic observables.
We take the opportunity to clarify some misconceptions in the literature concerning 
the identification of effective temperatures in out of equilibrium active systems. 
Finally, in Sec.~\ref{sec:open-pbms-dyn} we summarise our findings and we mention a 
number of future subjects of research. We also compare our results to other ones in the 
literature for diatomic active systems as well as particle systems.

\section{The dumbbell model}
\label{sec:model}

In this Section we present the definition of the model 
and we give a few details on the numerical methods used in our simulations.

\subsection{Equation of motion}

Each dumbbell is a diatomic molecule, 
consisting of two spheres of diameter $\sigma_{\rm d}$ and mass $m_{\rm d}$ linked together {\it via} a spring.
In the rigid limit in which vibrations are suppressed, the disks are kept at a fixed centre-to-centre distance 
equal to $\sigma_{\rm d}$, for a total number of $2N$ spheres. The two beads in each dumbbell
have a tail and head identity that they keep all along the evolution.

The Langevin equation of motion acts on the 
position of the centre of each bead, ${\mathbf r}_i$, and is given by
\begin{equation}
m_{\rm d} \ddot {\mathbf r}_i=-\gamma_{\rm d}\dot {\mathbf r}_i -{\boldsymbol \nabla}_i U+
{\mathbf F}_{\rm act}+
\sqrt{2k_BT\gamma_{\rm d}} \, {\boldsymbol \eta}_i(t) \; ,
\label{bd}
\end{equation}
where $i=1,\dots, 2N$ is the sphere index, $\gamma_{\rm d}$ is the friction coefficient, 
${\boldsymbol \nabla}_i=\partial_{\mathbf{r}_i}$, $T$ is the temperature of the thermal bath and $k_B$ is the Boltzmann constant. 

The last term in Eq.~(\ref{bd}) is proportional to ${\boldsymbol \eta}_i(t)$, a Gaussian white random noise satisfying
\begin{equation}
\begin{split}
\langle\eta_{ia}(t)\rangle & =0 \; , \\ 
 \langle\eta_{ia}(t_1)\eta_{jb}(t_2)\rangle & =\delta_{ij}\delta_{ab}\delta(t_1-t_2)
 \; , 
\end{split}
\end{equation}
where $a,b=1,2$ label the two spatial coordinates. Since the noises acting on the two beads 
that form a dumbbell are independent, the combined stochastic force can make the dumbbell
rotate.

\begin{figure}
\resizebox{\columnwidth}{!}{
  \includegraphics{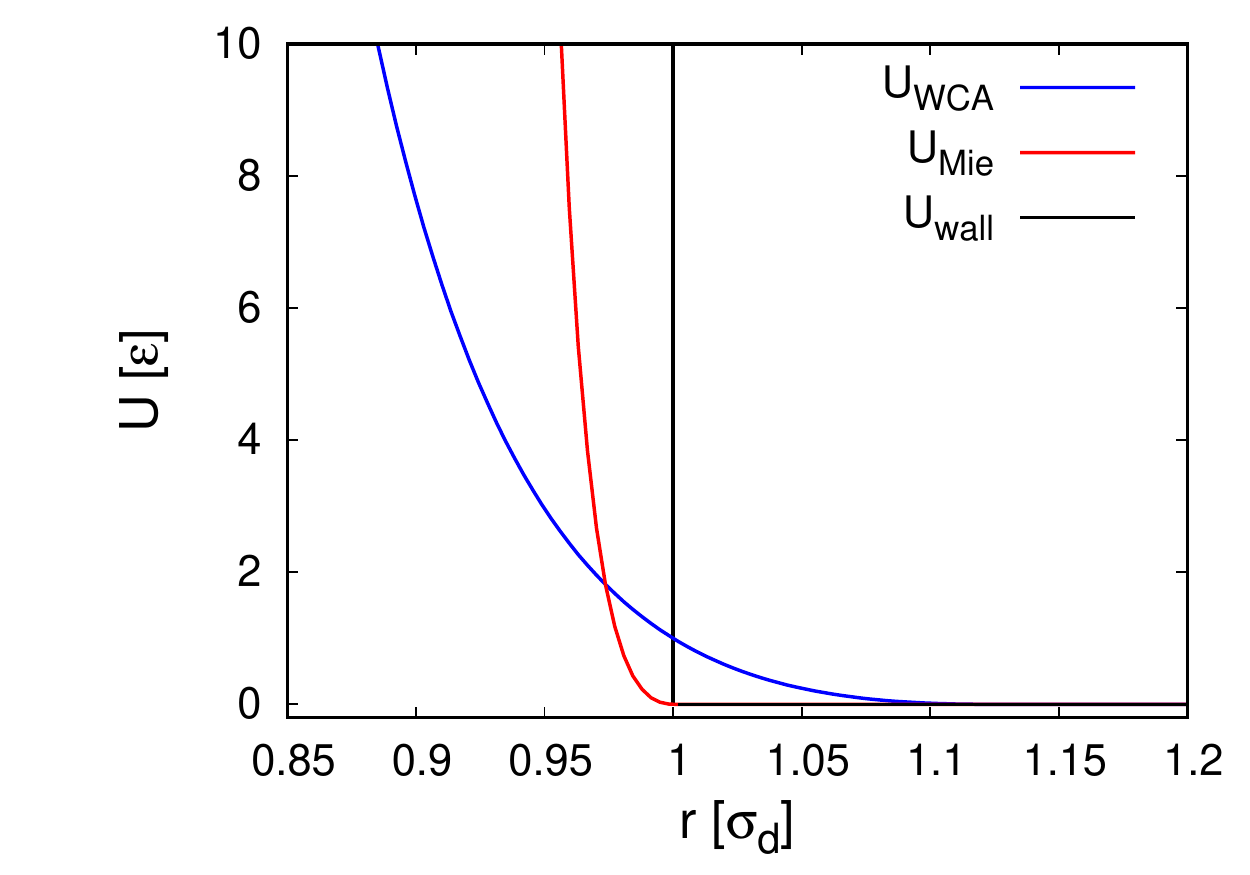}
  }
\caption{A comparison between the Weeks-Chandler-Andersen potential with a cutoff at its minimum $2^{1/6}\sigma_{\rm d}$ (blue curve), 
the Mie potential that we used and that is defined in Eq.~(\ref{eqmie}) (red curve) and a sharp wall potential (black line).
}
\label{fig:potential}
\end{figure}

The total internal potential energy of the system is
\begin{equation}
U= \sum_{i=0}^{2N}\sum_{j=0;j\neq i}^{2N} U_{\rm Mie}(|{\mathbf r}_i-{\mathbf r}_j|)  
\; ,
\end{equation}
with  $U_{\rm Mie}$~\cite{Mie1903} being a purely repulsive potential defined as
\begin{equation}
U_{\rm Mie}(r)=\left\{ 4\epsilon \left[ \left(\frac{\sigma}{r} \right)^{2n} - \left(\frac{\sigma}{r}\right)^{n} \right] +\epsilon \right\} \theta(2^{1/n}\sigma-r) 
\; ,
\label{eqmie}
\end{equation}
where $\sigma$ and $\epsilon$ are parameters setting length and energy scales of the potential.  
The functional form in Eq.~(\ref{eqmie}) is abruptly set to zero at its minimum, located at $r=2^{1/n}\sigma$, by 
the  Heaviside function $\theta$. We considered $2^{1/n}\sigma=\sigma_{\rm d}$, in order for the minimum to be equal to 
the disk diameter in such a way that the potential derivative be continuous in $\sigma_d$. 
A high $n=32$ value is chosen in order for $U_{\rm Mie}$ 
to be as close as possible to the hard-disk case without losing computational efficiency, 
see Fig.~\ref{fig:potential}. 
Note that Eq.~(\ref{bd}) also has implicit additional forces that keep the distance between 
the two spheres of each dumbbell constant. These are taken into account while performing the numerical integration, see Sec.~\ref{secnum}.
%The parameter $\epsilon$ sets the strength of the deterministic force.

The active force ${\mathbf F}_{\rm act}$ acts on the tail-to-head direction of each dumbbell and has constant modulus 
$F_{\rm act}$. Its direction changes following the individual dumbbell's orientation.

The initial positions and velocities of the dumbbells are specified in Sec.~\ref{subsec:initial}. We have made several choices 
in order to check that, after a sufficiently long transient, 
the dynamics reach a steady state in which the details of the initial configuration are forgotten.

The dimensionless control parameters are  the surface fraction covered by the beads, 
\begin{equation}
\phi=\frac{N\pi\sigma_{\rm d}^2}{2L^2}
\; , 
\end{equation}
and   the P\'eclet number, 
\begin{equation}
\mbox{Pe} = \frac{2F_{\rm act}\sigma_{\rm d}}{k_BT}
\; .
\end{equation}
The latter can be seen as the ratio between the advective transport  
of a fixed amount of material flowing  in unit time through a section of a characteristic  
length $\ell $  induced by the active force,
 $v_{\rm act} / \ell = F_{\rm act}/(\sigma_{\rm d} \gamma_{\rm d})$
and the diffusive transport of the same amount of material due to  thermal fluctuations, 
$D/\sigma^2_{\rm d}  = k_BT/(2\gamma_{\rm d} \sigma^2_{\rm d})$. It can also 
be interpreted as the ratio between the work done by the active force when translating the dumbbell by its 
%characteristic
 size, $2\sigma_{\rm d} F_{\rm act}$ over the thermal energy scale $k_BT$.
An active Reynolds number Re $=m_{\rm d}F_{\rm act}/(\gamma^2_{\rm d} \sigma_{\rm d})$ can be defined in analogy with
the hydrodynamic one. It is the ratio between the active force transport $ v_{\rm act}/ \ell$
% = \sigma_{\rm d} F_{\rm act}/\gamma_{\rm d}$ 
and the viscous transport term
%force
 estimated as $\gamma_{\rm d} /m_{\rm d}$ and takes very small values for the parameters chosen 
in all our numerical studies.  In practice, the active Reynolds number takes in our case values in the range  
Re $\in [0-0.05]$, and can be considered negligible, a choice that corresponds to the modelling of bacterial colonies, 
as mentioned in the introductory Section.
The value of the active force giving a speed value that is compatible with the one observed experimentally for bacteria (of the order of [20,50] $\mu m/s$) should imply in our simulations Pe values equal to several hundreds.

In the single dumbbell case Eq.~(\ref{bd}) can be solved analytically and a summary of the  motion 
in such extreme dilute limit will be given in Sec.~\ref{sec:single-dumbbell}. For a finite density system an 
analytical solution is out of reach and one has to resort to a numerical integration of the coupled equations. In the next Subsection we give some details on the numerical scheme implemented.

\subsection{Numerical Integration}
\label{secnum}

We fixed the number of particles in our system to  $N=128^2, \ 256^2, \ 512^2$ and, accordingly, 
we set $L^2=N\pi\sigma_{\rm d}^2/(2\phi)$ in order to have the desired packing fraction $\phi$.

All physical quantities are expressed in reduced 
units of the sphere's mass $m_{\rm d}$,  diameter $\sigma_{\rm d}$ and potential energy $\epsilon$~\cite{allen}. 
The time unit is the standard Lennard-Jones time 
$\tau_{LJ}=\sigma_{\rm d} (m_{\rm d}/\epsilon)^{1/2}$.
Other important simulation parameters that we used, in reduced units, are 
$\gamma_{\rm d}=10$, $k_BT=0.05$ and we set
$k_B=1$. The large $\gamma_{\rm d}$ assures the over-damped limit. 
Typical simulations took between $10^5$ and $10^6$ simulation time units (MDs), where longer runs are 
required to ensure stationarity in the co-existence region~\cite{Cugliandolo2017}. 
On average each simulation lasting $5 \times 10^5$ MDs was run on 16 processors for a total of 100 hours for each cpu. 

We used a velocity Verlet algorithm that solves  Newton's equations of motion, plus additional force terms for the Langevin-type  thermostat, 
to numerically integrate the stochastic evolution equation. 
We kept the  bonds  rigid with the help of the RATTLE scheme~\cite{rattle}. 
This is equivalent to considering an additional force in Eq.~(\ref{bd}), that  takes into account the holonomic constraints. 
The time-step choice is related to the force  
exerted during the simulation. We adapted it to enforce numerical stability. In this paper, for systems at Pe $\leq 10$ we used a 
time-step of $0.005$, while for Pe = 20 and Pe = 40 we used a time-step equal to $0.002$ and, finally,  for Pe = 100 and Pe = 200 the time-step was 
reduced to $0.001$.

In order to efficiently parallelise the numerical computation
we used the open source software Large-scale Atomic/Molecular Massively Parallel Simulator \linebreak (LAMMPS), available 
at {\tt github.com/lammps}~\cite{plimpton1995fast}.  

In addition,  we also studied the behaviour of the system using the 
Weeks-Chandler-Andersen (WCA) potential between the disks, that corresponds to $n=6$ in the potential $U(r)$ 
and a truncation at $r_c=2^{1/6}\, \sigma_{\rm d}$, see Fig.~\ref{fig:potential}. 
The dumbbells were still taken to be rigid with a distance $\sigma_{\rm d}$ between the centres of their beads. 
We will briefly mention the results found with this other potential in the text.

\section{Single dumbbell limit}
\label{sec:single-dumbbell}

The analytic solution to the Langevin equation regulating the dynamics of a single active
dumbbell was given in Ref.~\cite{Suma14b,Suma14c}. We do not repeat the derivation here but we simple 
summarise some features that will be relevant to identify important time scales and 
understand the dynamics of the interacting problem.

The individual dumbbell undergoes centre of mass displacement and rotational motion, the 
latter being due to the random noises that act independently on the two disks.

The equation ruling the centre of mass motion is a Langevin equation for a point-like particle with 
mass $2m_{\rm d}$, under the force $2F_{\rm act}$ and in contact with a bath with friction coefficient
$2\gamma_{\rm d}$ and temperature $T$.

\begin{figure}
\resizebox{\columnwidth}{!}{
  \includegraphics{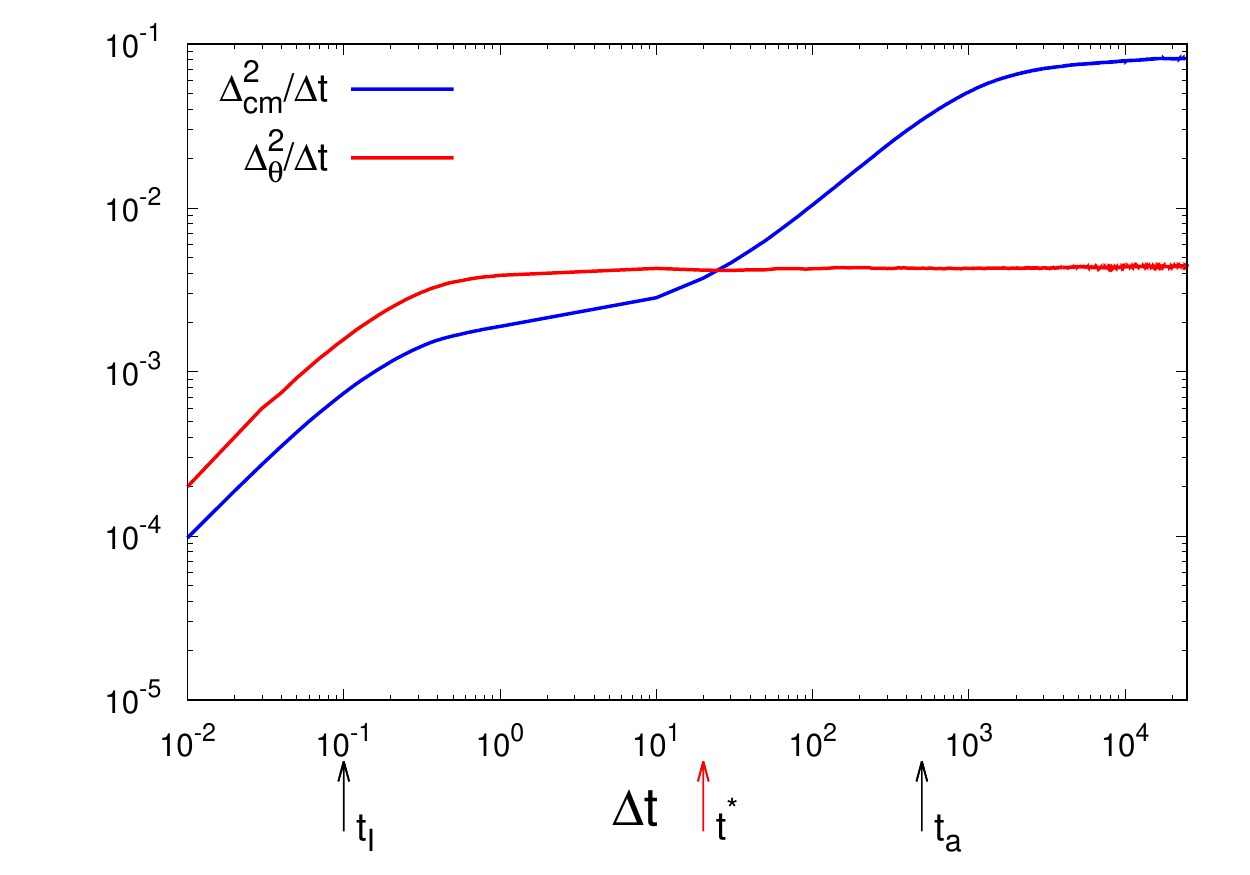}
  }
\caption{The single dumbbell centre of mass mean-square displacement  (MSD) divided by time delay, 
$\Delta^2_{\rm cm}/\Delta t$, as a function of $\Delta t$ is shown with a blue solid line.  The figure corresponds to   Pe = 20, with 
$F_{\rm act}=0.1$ and $T=0.01$.  The four regimes, ballistic, diffusive, ballistic and 
diffusive, can be seen in succession. The crossover times, indicated with arrows, are $t_I=0.1$, $t^*=20$ and $t_a=500$ MDs, 
as defined in the text. The angular MSD divided by time delay,
$\Delta^2_\theta/\Delta t$, is also shown with a red solid line and, for a single molecule, it has only two time scales, separated by $t_I$.
}
\label{fig:sketch-MSD}
\end{figure}

Four time regimes can be identified from the time-delay $\Delta t$ dependence of the 
mean-square displacement (MSD) of the centre of mass
\begin{equation}
\Delta^2_{\rm cm}(\Delta t) =  \langle ({\bold r}_{\rm cm}(t+\Delta t) -  {\bold r}_{\rm cm}(t))^2 \rangle
\; . 
\label{eq:MSD-CM-single}
\end{equation} 
Here and in what follows the angular brackets denote an average over independent
thermal noise histories and initial conditions.
The four regimes, shown in Fig.~\ref{fig:sketch-MSD} with a blue solid line,  are:
\begin{itemize}
\item
I. A ballistic regime at very short time differences, $\Delta t \ll t_I = 2m_{\rm d}/(2\gamma_{\rm d})$,
with velocity $v_{\rm cm} = k_BT/(2m_{\rm d})$. The factors 2 in the mass and the friction coefficient
reflect the fact that the dumbbell is made of two beads. The standard inertia time $t_I$ controls the crossover towards the 
next regime. In the over-damped limit that we will consider in the finite density case, $t_I$ is very short, $t_I \simeq 10^{-1}$ and
this regime is practically not observed in the numerical simulations.  
\item
II. A diffusive regime common to the one for the single passive dumbbell, with 
diffusion constant, $D=1/(2d) \lim_{\Delta t\to\infty} {\rm d}\Delta^2(\Delta t)/ {\rm d}t$ that takes 
the thermal \linebreak 
value $D = k_BT/(2\gamma_{\rm d})$. This 
regime lasts until a crossover dictated by the strength of the active force 
$t^* \propto t_a/\mbox{Pe}$ with $t_a=D_R^{-1} = \gamma_{\rm d} \sigma_{\rm d}^2 /(2k_BT)$, and $D_R$ the rotational diffusion constant, 
see below. A sufficiently strong active 
force can erase  this regime.
\item
III. A second ballistic regime due to the activity that  
lasts until $t_a$,
%. From $ {\rm d}\Delta^2_{cm}(\Delta t)/ {\rm d}\Delta t$ one extracts a ``velocity'' 
with a velocity $v \propto F_{\rm act}/\gamma_{\rm d}$.
\item
IV. A final diffusive regime with diffusion constant $D_A=k_BT (1+\mbox{Pe}^2/8)/(2 \gamma_{\rm d})$, that 
increases as Pe$^2$ for large active forces.
\end{itemize}
Obviously, if the parameters are such that the time scales $t_I$, $t^*$ and $t_a$ are not well separated, the regimes 
overlap.

Meanwhile the rotational motion can also be studied by following the angular MSD
\begin{equation}
\Delta_\theta^2(\Delta t) =  \langle (\theta(t+\Delta t) -  \theta(t))^2 \rangle
\; . 
\label{eq:MSD-theta-single}
\end{equation} 
In the {\it stiff} case in which the molecular vibrations are completely frozen,
the stochastic equation for the angle can also be solved and the angular MSD  computed. In the single molecule case, 
this quantity has less interesting features in the sense that there are only two time-regimes with a crossover from an initial ballistic one 
 to diffusive motion with diffusion constant $D_R=2k_BT/(\gamma_{\rm d} \sigma_{\rm d}^2)$ at $t_I$.
The rotational motion of a single dumbbell is independent of the active force.

The various time regimes described above also manifest in the probability distributions of the single 
run mean-square and angular displacements of a single dumbbell (not averaged over the noise). The results concerning these
quantities can also be found in~\cite{Suma14b,Suma14c}. We will come back to them in Sec.~\ref{sec:open-pbms-dyn}
where we will discuss the finite density effects on the dynamics of the many-body system.

As is well known~\cite{cugl:review,cugl-mossa1,cugl-kur-pel,wang,Jop08,greinert2006,LevisBerthier15}, 
under certain conditions (slow dynamics, small entropy production, etc.) the long term dynamics of some non-equilibrium systems
(glasses, dense sheared liquids, weakly shaken dense granular matter)
can be described (at least partially) 
 in terms of an effective temperature. One way to identify this temperature is to calculate the 
deviations from the fluctuation dissipation theorem (FDT) linking the integrated linear response 
\begin{equation}
\chi(\Delta t)= \frac{1}{d} \sum_{a=1}^d \int_t^{t+\Delta t} {\rm d}t' \; 
\left. 
\frac{\delta \langle r^a_{\rm cm}(t+\Delta t)\rangle}{\delta h^a(t')} \right|_{h=0}
\end{equation} 
to a perturbation ${\bold h}$ applied, in this case, to the centre of mass of the dumbbell,
and the mean-square displacement $\Delta^2_{\rm cm}(\Delta t)$ of the same observable, that is to say, the centre of mass of the 
dumbbell here. Using the expression of the FDT in the equilibrium form, 
\begin{equation}
2k_B T \chi(\Delta t) = \Delta^2_{\rm cm}(\Delta t)
\; , 
\end{equation}
one can extract the effective temperature by replacing $T$ with $T_{\rm eff}(\Delta t)$ in the formula above. 
One has, though, to carefully keep in mind 
that the dynamics take place in
various  time-regimes (see the four items listed above) 
and that the effective temperature may 
take different values in each of them or even have a meaning only in some of them~\cite{cugl:review}.

In Ref.~\cite{Suma14b} we also computed the effective temperature thus defined of a single active dumbbell and we 
found that in the last diffusive regime called IV, and defined by $\Delta t \gg t_a$,  
it takes the form
\begin{equation}
k_B T_{\rm eff} = k_BT \left( 1+ \mbox{Pe}^2/8 \right)
\; . 
\label{eq_teff}
\end{equation}
This effective temperature characterises the dynamics at long time-delays and should not be confused 
with, {\it e.g.}, the one extracted from the velocity fluctuations, $T_{\rm kin}$,  that can only access the instantaneous 
properties (high frequency regime) of the system. For example, for the single dumbbell the kinetic temperature extracted 
from the averaged kinetic energy and the equipartition assumption is equal to 
\begin{equation}
k_B T_{\rm kin} = \langle E_{\rm kin} \rangle = k_BT \ \left[1+ \frac{m_{\rm d}k_BT}{(2\gamma_{\rm d} \sigma_{\rm d})^2} \, \mbox{Pe}^2 \right]
\label{eq:Ekin}
\end{equation} 
and,
although the difference with $T_{\rm eff}$  
is just in the pre-factor of the term proportional to Pe$^2$, it is clear that the 
dependence on all the other parameters in the model is not the same. The conceptual and quantitative 
difference between $T_{\rm eff}$ and $T_{\rm kin}$ is well-known in the field of glasses and it will
manifest more clearly when dealing with many-body active dumbbells in interaction. We will come back 
to the interpretation of the effective temperature by the end of this paper, where we will discuss its 
properties in the many-body active system.

\section{Structure}
\label{sec:structure}

We first present the results of our study of the structural properties of the system.

\subsection{Observables}
\label{sec:observables}

In this Subsection we list and explain the definition of  the various observables that we used to quantify the 
degree of order in the dumbbell system and to identify the  different phases.

\subsubsection{Local density fluctuations}

We used two methods to compute
 the {\it local densities} $\phi_j$ and we checked that they give equivalent results. 

With the first method, for each bead, we first estimate the local density as the ratio between its surface  
and the area of its Voronoi region $A_i^{\rm Vor}$. We next coarse-grain this value by averaging the single-bead densities over 
a disk with radius $R$ around the $j$th bead, 
\begin{equation}
\phi_j = \frac{1}{n_R^{(j)}} \sum_{i \in S_R^{(j)}} \frac{\pi \sigma_{\rm d}^2}{4 A_i^{\rm Vor}}
\; ,
\end{equation}
where $n_R^{(j)}$ and $S_R^{(j)}$ are the number of particles and the spherical disk with radius $R$ around the $j$th bead, respectively.
We took, typically, $R = 20 \, \sigma_{\rm d}$, and we found equivalent results using 
radii in  the interval $[10, 50] \, \sigma_{\rm d}$. At Pe $>$ 10 we also used  
$R = 5 \,\sigma_{\rm d}$.  For a visual inspection of the density fluctuations in real space, 
each Voronoi region is then painted with the colour that corresponds to its coarse-grained local density, 
using a heat map with the usual convention: denser in red, looser in blue. 

With the second method,
we constructed a square grid on the simulation box. For each point in the grid we calculated a coarse grained 
local density over a circle of given radius $R$. We finally assigned this density value to the grid point.

The data stemming from the use of  both methods are collected and 
pdfs of the local densities thus obtained are built. Some of them are 
shown below.

\subsubsection{Orientational order}

With the aim of exhibiting orientational order, we computed the {\it local hexatic order parameter}
\begin{equation}
 \psi_{6j}=\frac{1}{N^j_{\rm nn}}\sum_{k=1}^{N^j_{\rm nn}}e^{6{\rm i}\theta_{jk}},
 \label{loc_hex_par_def}
\end{equation}
where $N_{\rm nn}^j$ is the number of nearest neighbours 
of bead $j$ found with a Voronoi tessellation algorithm~\cite{voro++} and 
$\theta_{jk}$ is the angle between the segment that connects $j$ with its neighbour $k$ 
and the $x$ axis. For beads regularly 
placed on the vertices of a triangular lattice, each site has six nearest-neighbours, all the angles are integer multiples of $2\pi/6$, $\theta_{jk}=2k\pi/6$, 
and $\psi_{6j} =1$. Deviations from $1$ indicate deviations from perfect orientational ordering.

We visualised the local values of $\psi_{6j}$ as proposed in Ref.~\cite{BeKr11}:
first, we projected the complex local values $\psi_{6j}$ onto the direction of their space average. 
Next, each bead was painted according to this normalised projection.
Zones with orientational order have uniform colour, whatever it is. The majority ordering is 
always coloured in dark red and the hierarchy follows the scale shown at the extreme right of the panels in the figures. 
Due to the six-fold symmetry of the ordered state, blue regions are hexatically ordered along a lattice which is rotated by $\pi/2$ with respect to the one of 
the dark red regions. Green spots, which correspond to zero in the colour code, can represent either particles with disordered neighbours or 
ordered regions rotated by $\pi/4$ from the red ones. Therefore, while the greenish regions surronding clusters are disordered, also macroscopic green patches appear, which are hexatically ordered.

With the local hexatic order parameter we can compute correlation functions
\begin{equation}
g_6(r) = \frac{\left. \langle \psi^*_{6j} \psi_{6k}\rangle\right|_{|\vec r_j - \vec r_k|=r}}{\langle |\psi_{6j}|^2\rangle}
\; .
\label{eq:hec_corr_func}
\end{equation}
The conventional HNY scenario for passive systems predicts that this correlation function should approach a constant in the
solid phase, it should vanish as a power law in the hexatic phase, and it should decay exponentially 
in the liquid phase.

We also considered the modulus of the average per particle and the average per particle of the modulus
of the local hexatic order parameter,
\begin{eqnarray}
2N \, \psi_6 \equiv  \Big{|} \sum_{j=1}^{2N} \psi_{6j} \Big{|}
\; , 
\qquad
2N \, \Gamma_6 \equiv  \sum_{j=1}^{2N} |\psi_{6j}|
\label{eq:defs}
\end{eqnarray}
and their variation with the two control parameters $\phi$ and Pe.

\subsubsection{Positional order}

Positional order is put to the test by the usual two-point correlation functions
\begin{equation}
C_{\bold{q_0}} (r) = \langle e^{{\rm i} \bold{q_0} \cdot ({\bold r}_i - {\bold r}_j)}\rangle
\label{eq:position-correlation}
\end{equation}
with $r = |{\bf r}_i - {\bf r}_j|$ and $\bold{q_0}$
the wave vector  that corresponds to the maximum value of the first diffraction peak of the structure factor:
 \begin{equation}
S(q) = \frac{1}{2N}  \sum_{i=1}^{2N} \sum_{j=1}^{2N} \langle e^{{\rm i} {\mathbf q} \cdot ({\mathbf r}_i-{\mathbf r}_j)} \rangle
\; .
\label{eq:structure-factor}
\end{equation}

In the arguments of $C$ and $S$ we have already used the translational invariance and isotropy of the system (on average) and 
we only wrote the absolute values of the distance and wave vector.
The quasi-long range positional order in the solid phase should be evidenced by the algebraic decay of the positional correlation 
function. Instead,  in the hexatic and liquid phases the decay of $C_{\bold{q_0}} (r)$ should be exponential. 
Note also that the formulas for $C$ and $S$  explicitly need  the  use of the disks positions, and not the dumbbells centres of mass to observe hexatic and positional orders. In fact, while the  disks constituting  the dumbbells  pack at high density on a triangular lattice, the dumbbells centres of mass form non-periodic structural arrangements when packed~\cite{Frenkel91}.  

\subsubsection{Polarisation}
\label{subsec:polarization}

In order to investigate the alignment of head-tail dumbbell orientations inside the clusters, which is absent 
in the passive case but expected to depend on the strength of the active force,
we computed the 
\emph{local polarisation field} ${\mathbfcal{P}}$ 
and we evaluated its probability distributions. For each dumbbell in the dense hexatically ordered phase we considered a unit vector ${\vec{p}_i}$ in 
the tail-to-head direction. We constructed a square grid on the simulation box and for each point on the grid we calculated the spatial component of a 
coarse grained local polarisation by averaging over the values that this quantity takes for all dumbbells within
a disk with a previously chosen radius $R$. We assigned the averaged outcome to the 
grid point. Normalisation is such that a perfect polar order in each cell of the grid would give a vector of unit magnitude.
We used different radii, ranging from ${R=5 \, \sigma_d}$ to ${R=25 \, \sigma_d}$, finding qualitatively similar results. 
To visualise the local values of ${\mathbfcal{P}}$, we evaluated its modulus and we coloured each zone according to its magnitude.

\subsection{Numerical results}
\label{sec:numerics}

In this Section we sum up some of the results that we obtained in previous studies of the 
interacting active dumbbell system~\cite{Suma13,Suma14,Suma14b,Suma14c,Suma15a,Suma16,Cugliandolo2017} 
and we present data for quantities that we had not 
studied before and complement our analysis. We chose to show data for parameters on 
curves in the phase diagram that lie within the region with co-existence and on which a fixed 
area proportion of the two phases, 25-75\% and 50-50\%, is maintained. (We will always give the percentage of dense phase first.) 
The selected points in the phase diagram 
for which we show more results
are taken to be at the parameter values Pe = 0, 10, 20, 40, 100, 200  and 
the corresponding $\phi$s to ensure the chosen repartition of phases.

\subsubsection{Initial states}
\label{subsec:initial}

In all our simulations we used three kinds of starting configurations that we call `striped', 
`random' and `hexatic-ordered'. These configuration were obtained as follows. 

{\it Random initial conditions.}
We placed the dumbbells at random positions, with random orientations, 
in continuous space. At large $\phi$ these states have, very likely, overlapping 
dumbbells. Consequently,  these  configurations can have a very high energy.  
For the parameters used, the maximum overlap that we accepted is $0.75 \, \sigma_{\rm d}$. Therefore, 
we released the excess energy by letting the system 
equilibrate with a softer Lennard-Jones potential and a smaller time-step equal to $0.001$. After this procedure, 
convenient starting configurations as the one shown in the first figure in the Supplemental Material in~\cite{Cugliandolo2017} (a), 
were found.

{\it Striped states.}
We placed the dumbbells in a closed-packing triangular lattice, by placing the dumbbells in sequence starting from the bottom-left end, say.
This construction is followed until the dumbbell number satisfies the required global density. An empty slab is therefore left 
in order to have the right global density. 
The orientational order in the configuration is then randomised with Monte Carlo moves that 
take two adjacent dumbbells and exchange their bonds. In this way, the molecules are
still placed in a crystalline configuration but their directions are not ordered.

{\it \textcolor{black}{Hexatic} ordered states.} 
We took a crystalline configuration in equilibrium with \textcolor{black}{hexatic} order like the one depicted,
with packing fraction just above the co-existence region at Pe $=0$, obtained by equilibrating the system 
starting from a  striped configuration with this packing fraction.   
We then expanded (or contracted) the configuration by multiplying the coordinates with a convenient factor $\alpha$.
Accordingly,  the system's linear size is now rescaled to $\alpha \, L$, which is at most $20\, \sigma_{\rm d}$ larger (or smaller) than the original one, 
depending on the chosen density that we want to simulate. Finally we applied a short equilibration run with a smaller time-step equal to $0.001$.  
For the densities considered, the rescaling is small enough to preserve the initial \textcolor{black}{hexatic} ordering 
without blowing up the simulation.  

The long-time limit of the evolution of initial states of these three  types are statistically equivalent
for all parameters used. We will not show here results for all these cases, as more details on how the initial configurations are 
forgotten dynamically can be found in~\cite{Cugliandolo2017}.

\subsubsection{Phase diagram}

Following initial studies in, {\it e.g.}, \cite{Suma13,Suma14,valeriani2011colloids} 
that focused on the boundary enclosing MIPS at high Pe,  
in~\cite{Cugliandolo2017} we presented the full phase diagram of the active dumbbell model.
The 
main point made in this paper is that

\begin{itemize}
\item[$\bullet$]
co-existence between a dense and a dilute phase extends all the way to Pe = 0, with no ``critical'' ending point at a non-zero Pe, in a scenario different from what has been stated in the literature so far.
\end{itemize}

The dilute phase has no order and behaves as an active liquid or gas while the dense phase has orientational order and it is therefore an active hexatic phase.
These results made contact with the recent advances in the understanding of two-dimensional melting of passive disk systems in interaction: namely, the first order kind of the transition between liquid and hexatic phases and the ensuing co-existence between the two~\cite{Weber1995a,BeKrWi09,BeKr11,Engel13,KaKr15}.
Here, we extend the results found for Pe $\le$ 40 in~\cite{Cugliandolo2017} to Pe $\leq$ 200.

The phase diagram is displayed in Fig.~\ref{fig:phase_diagram} with the grey level indicating the percentage of area occupied by the dense hexatically ordered phase. 
It is based on a detailed plan of simulations performed at the values of Pe mentioned above. 
The various coloured lines are curves of constant area covered by the dense phase. We explain in the next Subsection how we obtained this phase diagram  using different observables. Notice that since the disks forming the dumbbells are not completely hard, some overlap between them is possible and values of $\phi$ that are slightly larger than the close packing limit can be accessed in the 
simulation (recall that the close packing fraction of disks in two dimensions is achieved by a perfect triangular lattice and it amounts to $\phi_{\rm cp} \approx 0.91$).

\subsubsection{The local density and hexatic order}

One way to estimate the critical lines $\phi(\mbox{Pe})$ 
for the upper limit for the pure liquid phase (dotted black curve in Fig.~\ref{fig:phase_diagram}) 
and the lower limit for the pure solid phase (white dotted curve in the 
same plot), is to follow the evolution of the
probability distributions (pdf) of the local density and modulus of the hexatic order parameter, and search for the appearance and 
disappearance of the two peak structure in these functions~\cite{Cugliandolo2017}. 
 \begin{figure}[h!]
 \begin{center}
\resizebox{0.9\columnwidth}{!}{
  \includegraphics{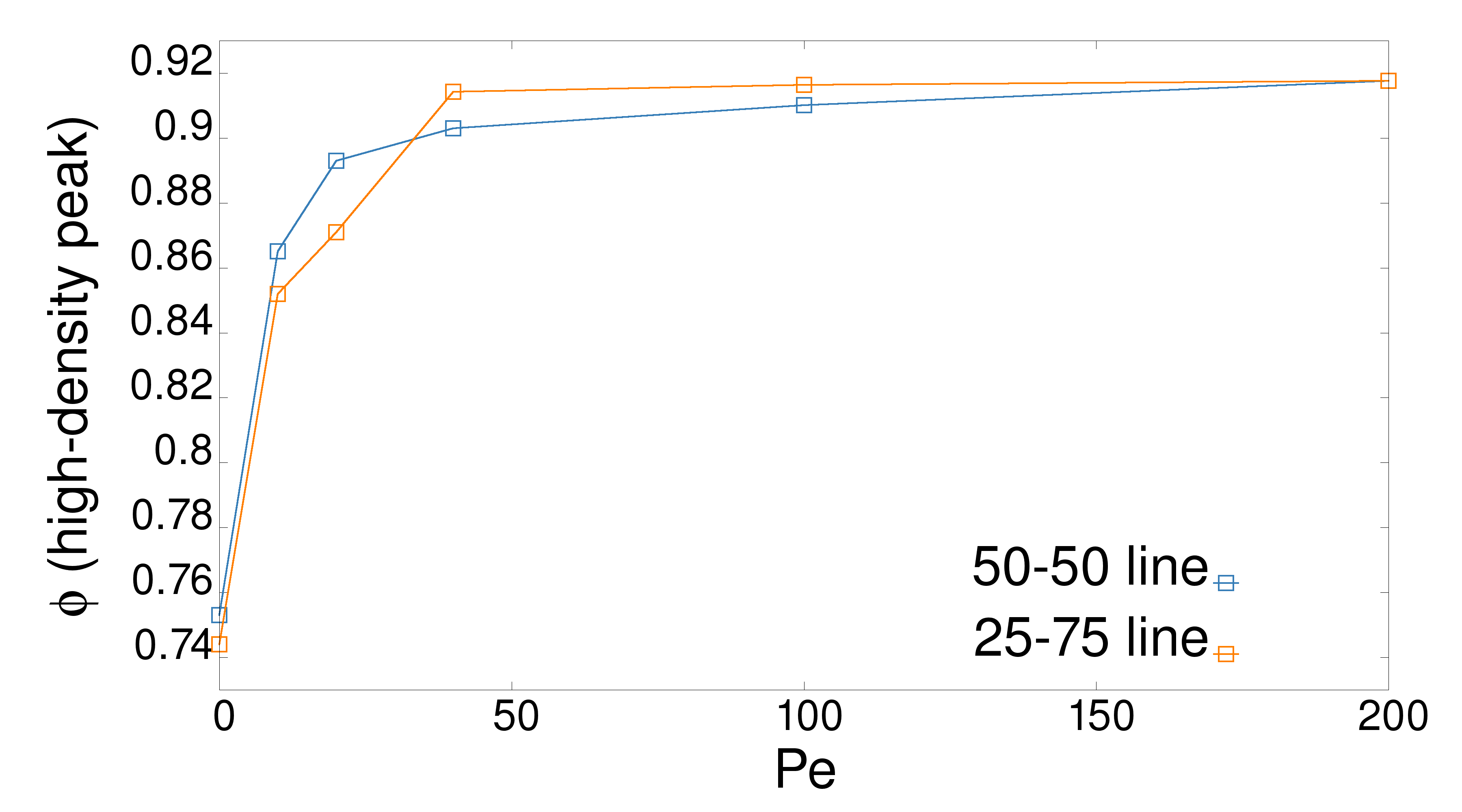}
}
\end{center}
\caption{Pe number dependence of the position ($\phi$ value) of the high-density peak of the bimodal probability 
density functions of local surface fraction (see  Figs.~\ref{fig:pdfs_25-75} and \ref{fig:pdfs_50-50}). The two curves correspond to the behaviour along the 50\%-50\% 
and 25\%-75\% area covered curves (here and in what follows the first percentage is the one covered by the dense phase).}
\label{fig:high-density-peaks}
\end{figure}

Figure~\ref{fig:high-density-peaks} represents the value of the local density
at which the high density peak of the pdf is located as a function of the P\'eclet number, 
following the curves of 50\%-50\% and 25\%-75\% repartition of dense and loose regions in the sample. At Pe = 0 the curves take a
non-vanishing value proving that there is co-existence in the passive limit. 
The growth with Pe is then monotonic,  the curves grow slowly, and they go a bit beyond the close packing limit
at large Pe due to the  fact that the potential is slightly soft. 

\begin{figure*}[h!]
\vspace{0.4cm}
$\;$ \hspace{0.25cm}  (a) \hspace{2.2cm} (b) \hspace{2.2cm} (c) \hspace{2.2cm} (d) \hspace{2.3cm} (e)  \hspace{2.2cm} (f) $\:$
\\
\resizebox{\textwidth}{!}{
  \includegraphics{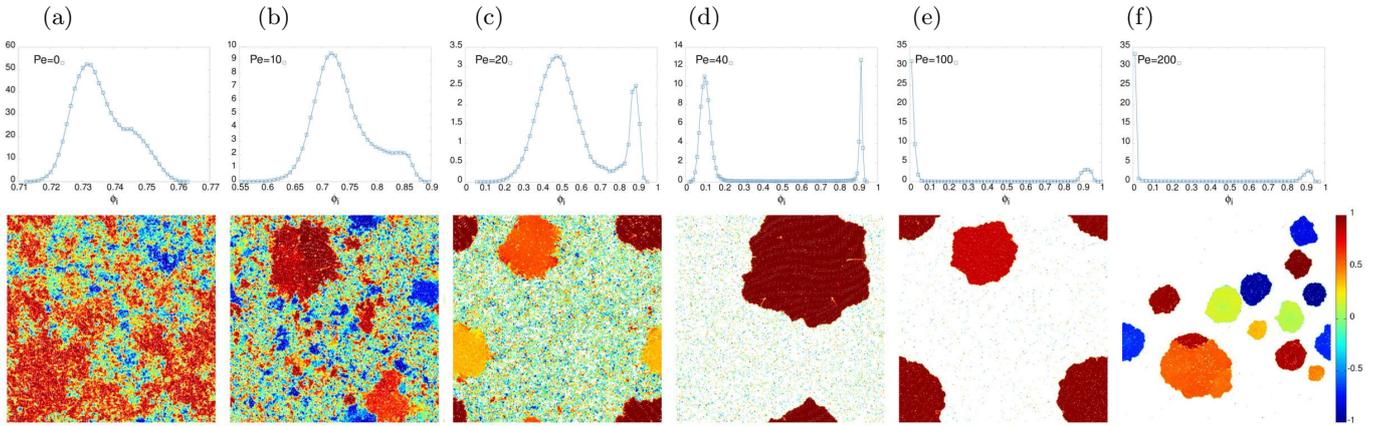}
}
\vspace{0.2cm}
\caption{Pdfs of local packing fraction (above) and instantaneous local hexatic order parameter maps (below) at different Pe numbers shown as labels in the upper plots,
 along the curve with an amount of $25$\% in area of the dense phase. The pdfs are evaluated by averaging over configurations 
 taken from at least five independent runs in the stationary state. On the second row we show some illustrative configurations taken after
 ${500000 \ \tau}$, when we can consider the systems in a stationary regime. As pointed out in the main text, 
 especially at Pe = 200  a single macroscopic hexatic phase is not formed due to the  
timescale of clusters diffusion, which rules aggregation, being much longer than the duration of our simulations. Even though in this case
we performed up to ten independent runs finding a similar behaviour, we do expect further coalescence, at still longer times.
The colour code of the snapshots, present at the right extreme,  is such that each bead is coloured  
according  to the local hexatic order parameter $\psi_{6j}$, defined in Eq.~(\ref{loc_hex_par_def}), projected onto the direction of its 
global average. For this reason, the 
colour appearing most often is the closest to one (dark red) and zones with the same uniform colour are associated to the same $\psi_{6j}$ value, 
and therefore have the same kind of orientational order. 
}
\label{fig:pdfs_25-75}
\end{figure*}
 \begin{figure*}[h!]
 \vspace{0.45cm}
 $\;$ \hspace{0.25cm}  (a) \hspace{2.2cm} (b) \hspace{2.2cm} (c) \hspace{2.2cm} (d) \hspace{2.3cm} (e)  \hspace{2.2cm} (f) $\:$
\\
\resizebox{\textwidth}{!}{
  \includegraphics{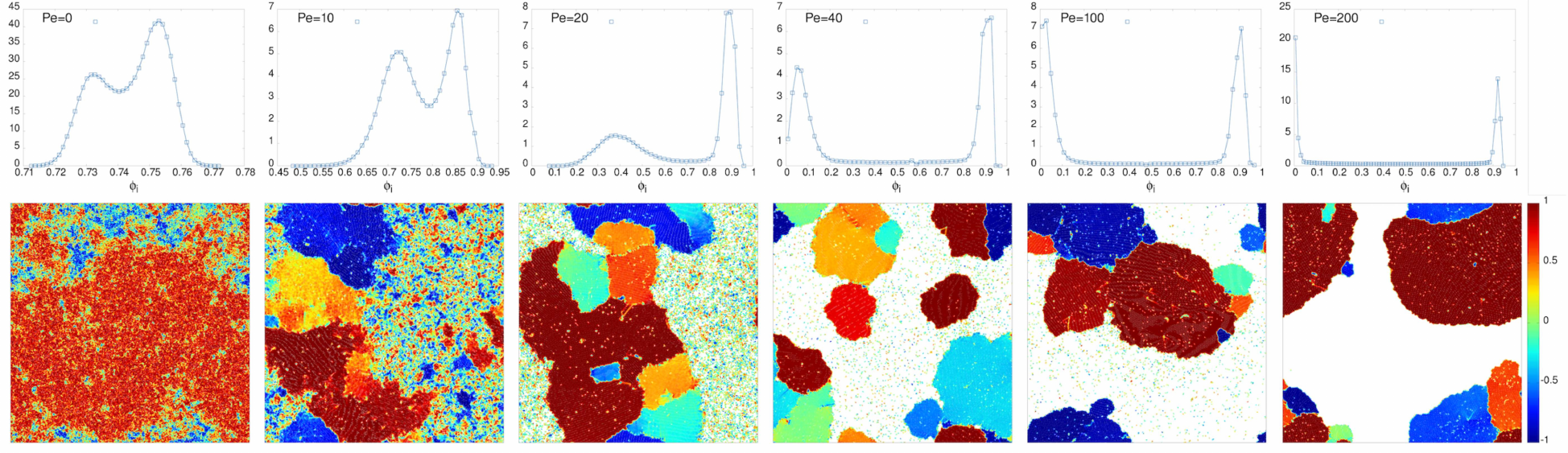}
}
  \vspace{0.2cm}
\caption{Pdfs of local packing fraction (above) and maps of local hexatic order parameter (below) at 
different Pe numbers along the curve with equal percentage in area of the dilute and dense phases. 
The presentation and colour code are the same as in Fig.~\ref{fig:pdfs_25-75} and they are explained in its caption. For these parameters, the 
total time used in the simulation was enough to obtain a single dense cluster (though polycrystalline from the orientational point of view) in 
all cases apart from (d).}
\label{fig:pdfs_50-50}
\end{figure*}

The data in Fig.~\ref{fig:high-density-peaks} were obtained from the analysis of the 
pdfs of $\phi_i$ shown in Figs.~\ref{fig:pdfs_25-75} and \ref{fig:pdfs_50-50}, together 
with snapshots of the configurations using a colour code
that represents the different hexatic orders. The data are taken along 
curves of constant proportion  25\%-75\%  (Fig.~\ref{fig:pdfs_25-75}) and  50\%-50\%   (Fig.~\ref{fig:pdfs_50-50}) 
of the two phases.  Note that the vertical scale of the different panels in the first row is not the same.
We see that 
\begin{itemize}
\item[$\bullet$]
as Pe increases,  the high density peak continuously moves towards
higher local densities and its weight increases while the low density peak moves in the opposite direction
and its weight decreases, leading to the results already shown in Fig.~\ref{fig:high-density-peaks}.
As far as density is concerned we do not find any discontinuity when moving towards higher activities 
in the coexistence region.
\end{itemize}
 
The local hexatic order parameter maps shown in the lower rows in Figs.~\ref{fig:pdfs_25-75} and Figs.~\ref{fig:pdfs_50-50} allow one to 
understand the mechanism whereby phase separation takes place in the active problem and the reason why the region of the phase diagram with co-existence inflates as Pe increases. Indeed,
\begin{itemize}
\item[$\bullet$]
as the activity is turned on, some spatial regions get denser and acquire orientational order, 
leaving away disordered holes. Depending on the strength of the activity, 
this process allows the dumbbells in the ordered regions to pack in a single ordered domain, or in polycrystalline arrangements concerning the 
orientational order. 
\end{itemize} 
In the maps in Fig.~\ref{fig:pdfs_25-75} and \ref{fig:pdfs_50-50} the fluid disordered regions are greenish while the denser, orientationally 
ordered ones, take all the other colours. At high Pe values, for instance in the Pe = 200 case, a single macroscopic hexatic phase is not 
yet formed. At lower area percentage of the higher density phase this is due to the fact that the timescale of clusters diffusion, 
which rules aggregation, goes well beyond the duration of our simulations. 
Moreover, the defects between the various orientationally ordered domains in the same clusters, 
that can be seen  when  the two phases occur with equal area percentages, 
should also slowly heal with time, 
but our simulations do not reach those exceedingly long time-scales.

\subsubsection{Polarisation}
\label{subsec:polarisation_results}

\begin{figure*}[t!]
$\;$ \hspace{0.25cm}  (a) \hspace{2.2cm} (b) \hspace{2.2cm} (c) \hspace{2.2cm} (d) \hspace{2.3cm} (e)  \hspace{2.2cm} (f) $\:$
\\
\resizebox{\textwidth}{!}{
  \includegraphics{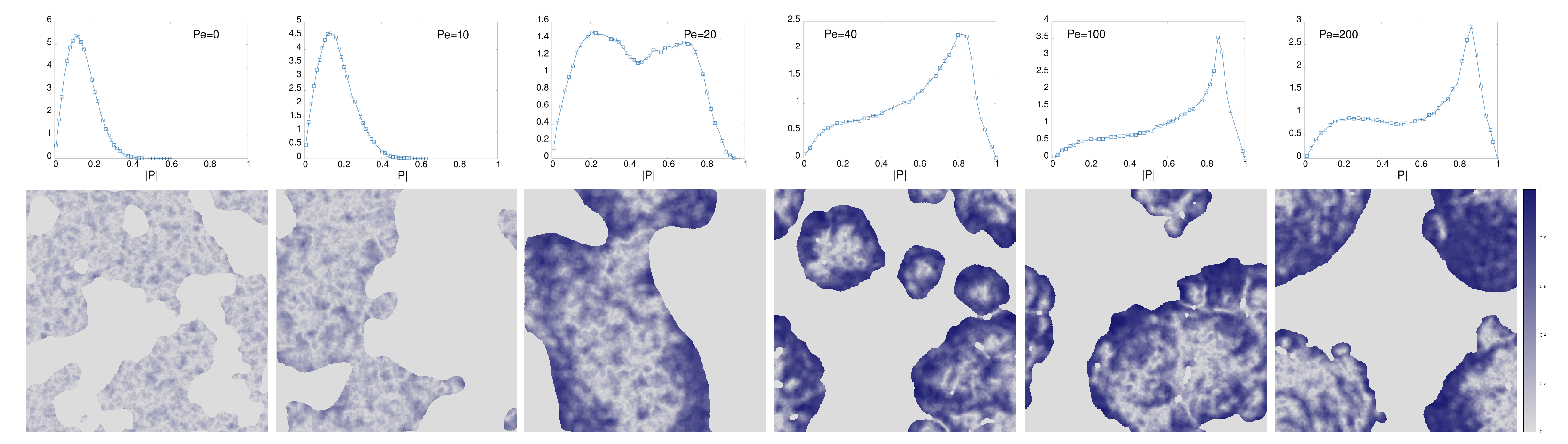}
} 
\vspace{0.05cm}
\caption{Pdfs of the local polarisation modulus $|{\mathbfcal{P}}|$  (above) and map of the local polarisation modulus (below)
both smoothed over regions with radius ${R=5 \, \sigma_{\rm d}}$  at different Pe number ranging from Pe = 0 
to Pe = 200 along the curve with $50$\%-50\% proportion of  hexatically ordered and liquid regions. The appearance in the pdfs of a second peak at 
high $|{\mathbfcal{P}} |$   and increasing Pe, signals that dumbbells prefer to be locally aligned in clusters, especially at the border. 
 The pdfs are evaluated by an average over stationary configurations taken from independent runs, as explained in the caption
 of Fig.~\ref{fig:pdfs_25-75}. For clarity, the snapshots of the local polarisation modulus are taken from the same run in Fig.~\ref{fig:pdfs_50-50}, 
 though at different times.
 }
\label{fig:pdfs_pol_50-50}
\end{figure*}
\begin{figure*}[t!]
\vspace{0.55cm}
$\;$ \hspace{0.25cm}  (a) \hspace{2.2cm} (b) \hspace{2.2cm} (c) \hspace{2.2cm} (d) \hspace{2.3cm} (e)  \hspace{2.2cm} (f) $\:$
\\
\resizebox{\textwidth}{!}{
  \includegraphics{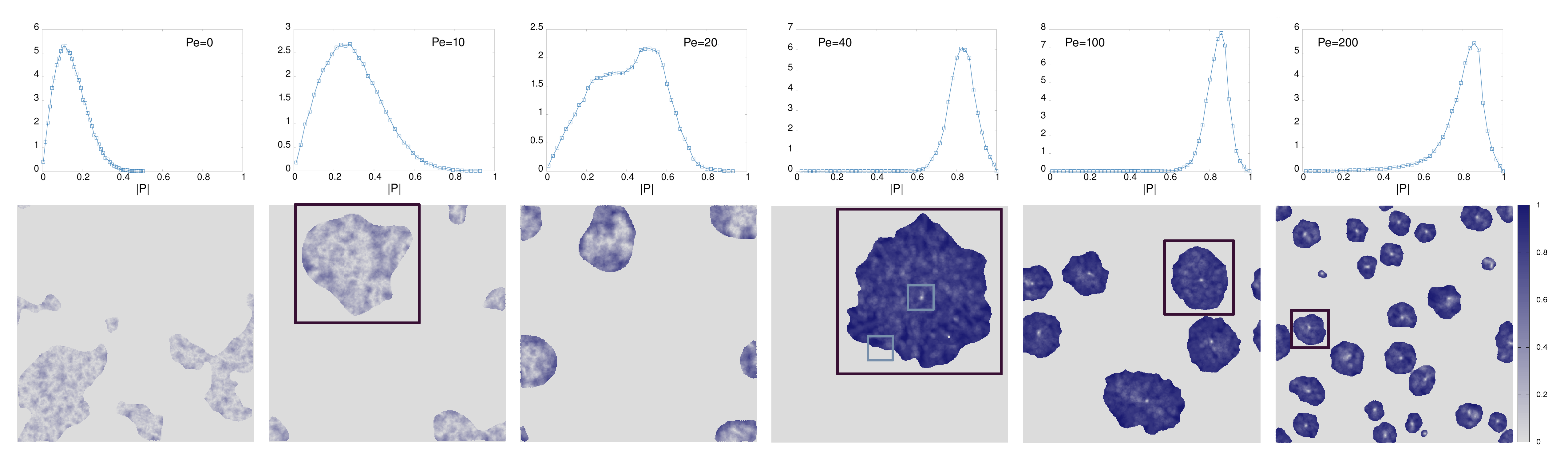}
}
%\vspace{0.1cm}
\caption{The same quantities as in Fig.~\ref{fig:pdfs_pol_50-50} 
here smoothed over regions with radius ${R=10 \, \sigma_{\rm d}}$ at the same Pe numbers as in the previous 
figure, now taken along the curve with an amount of $25$\% in area of the dense phase. Zooms over the clusters 
surrounded by black boxes in (b), (d), (e), (f) are shown in Fig.~\ref{fig:topology_polarization}, and over those surrounded by two cyan 
boxes at the centre and border of the cluster in (d) are shown in Fig.~\ref{fig:cluster_polarization}.
}
\label{fig:pdfs_pol_25-75}
\end{figure*}

We follow the evolution of the probability distributions of the local polarisation modulus  in the dense phase, 
defined in Sec.~\ref{subsec:polarization}, along the curves of constant area fraction 50\%-50\% (Fig~\ref{fig:pdfs_pol_50-50}) and  25\%-75\% 
 (Fig.~\ref{fig:pdfs_pol_25-75})  of the liquid and hexatic 
phase.

For low Pe values (Pe = 0, 10) the pdfs 
show a single peak at small values of the local polarisation modulus both on the 50\%-50\% and 25\%-75\% lines, 
revealing the absence of any polar order. This is also evident from the coarse-grained polarisation fields of the clusters, that look 
completely disordered regardless of being inside or at the border of clusters and dominated by low-polarisation values with sporadic negligible ordered 
parts.

\begin{figure*}[h!]
\vspace{0.55cm}
\begin{center}
(a) \hspace{3.9cm} (b) \hspace{3.9cm} (c) \hspace{3.9cm} (d) \hspace{3cm} $\;$
\\
  \begin{tabular}{cccc}
	\resizebox{0.5\columnwidth}{!}{ \includegraphics{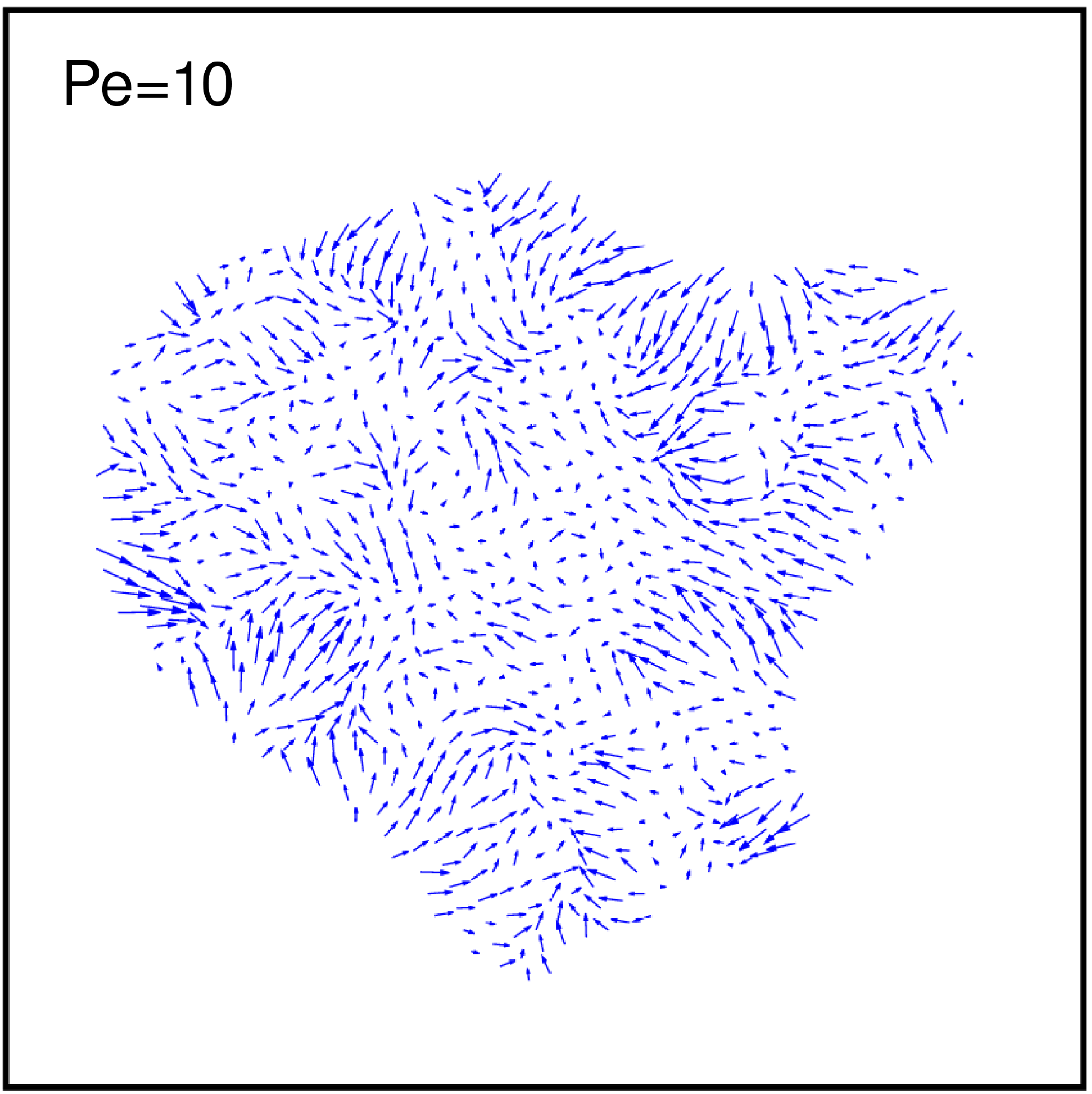}}
	\resizebox{0.5\columnwidth}{!}{ \includegraphics{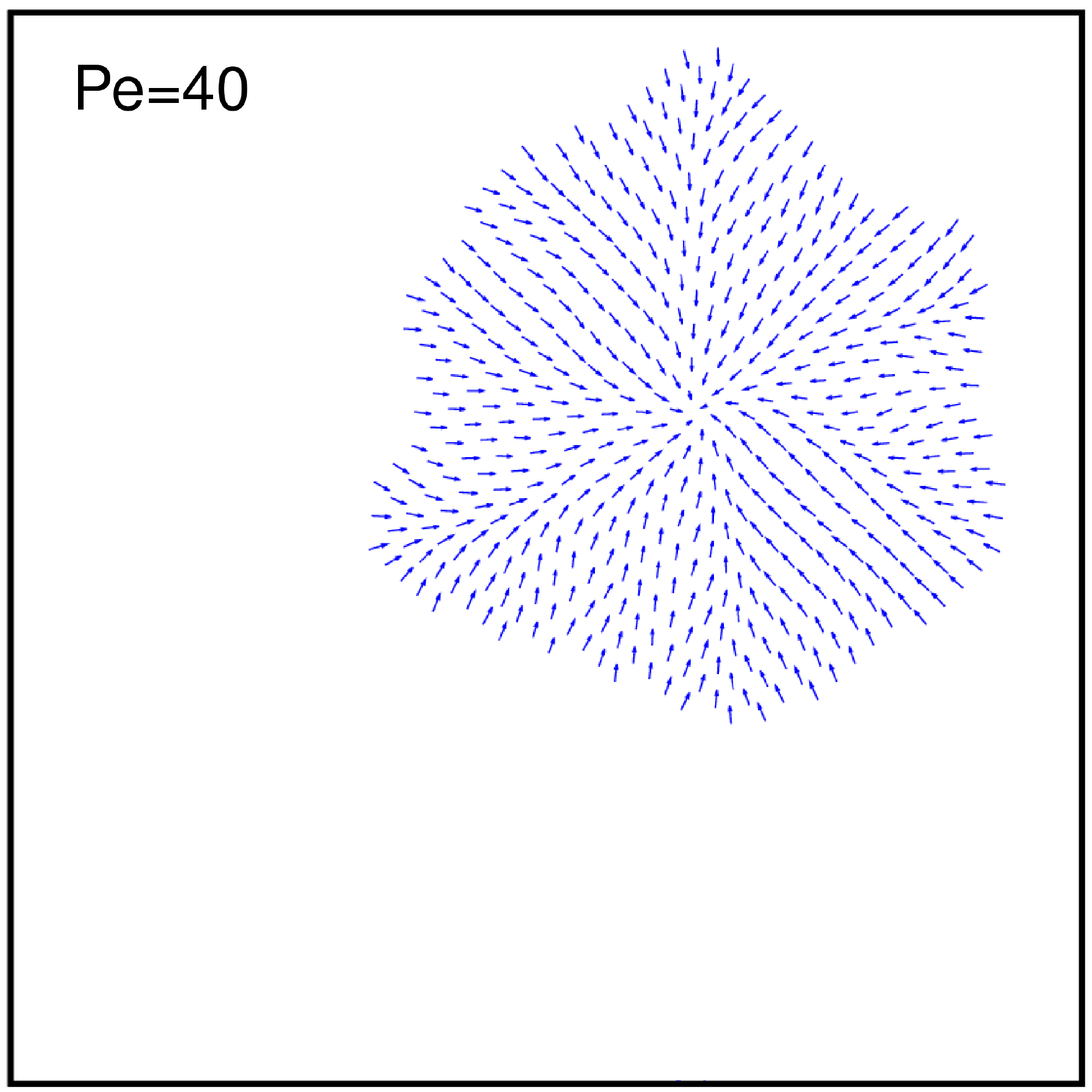}}
	\resizebox{0.5\columnwidth}{!}{ \includegraphics{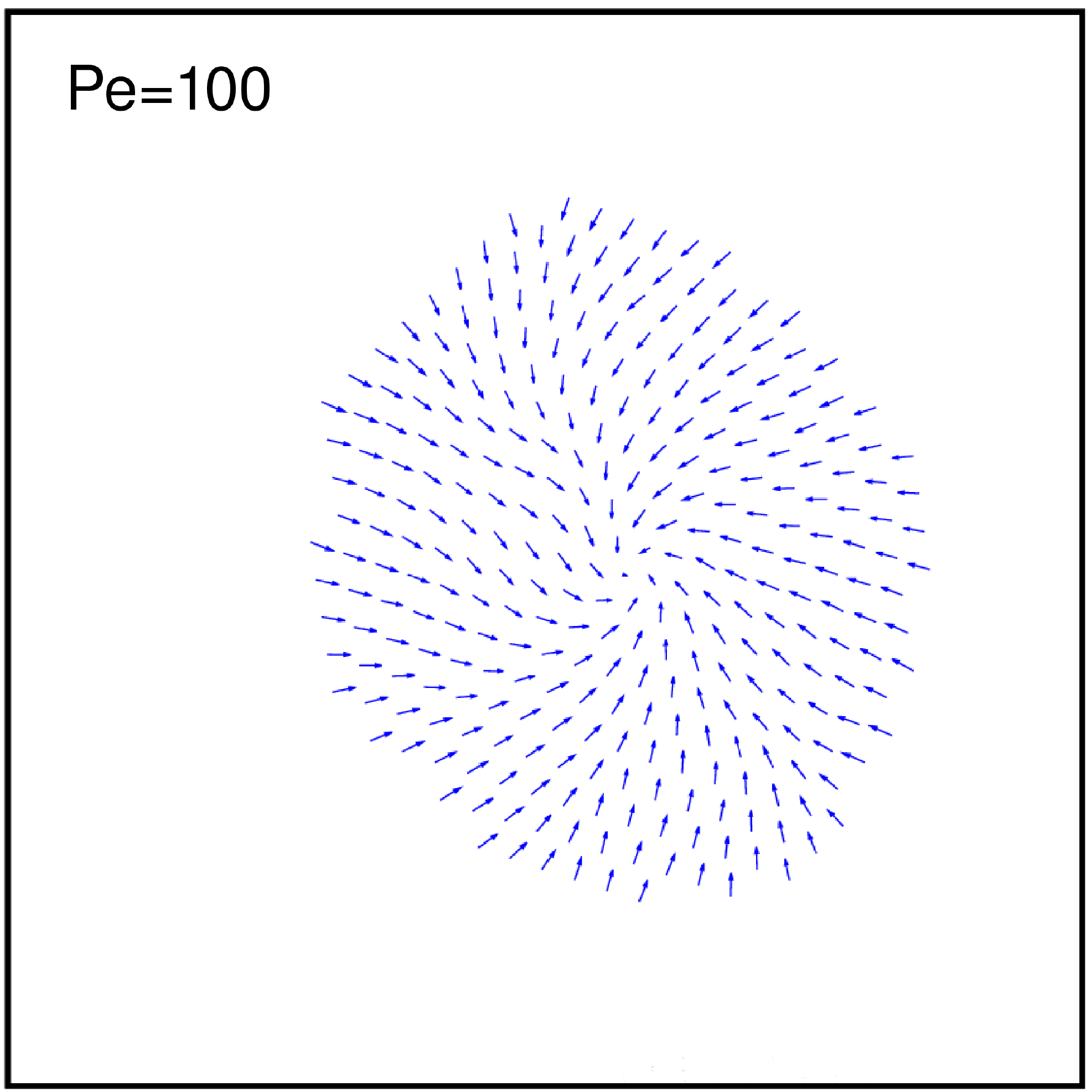}}
	\resizebox{0.5\columnwidth}{!}{ \includegraphics{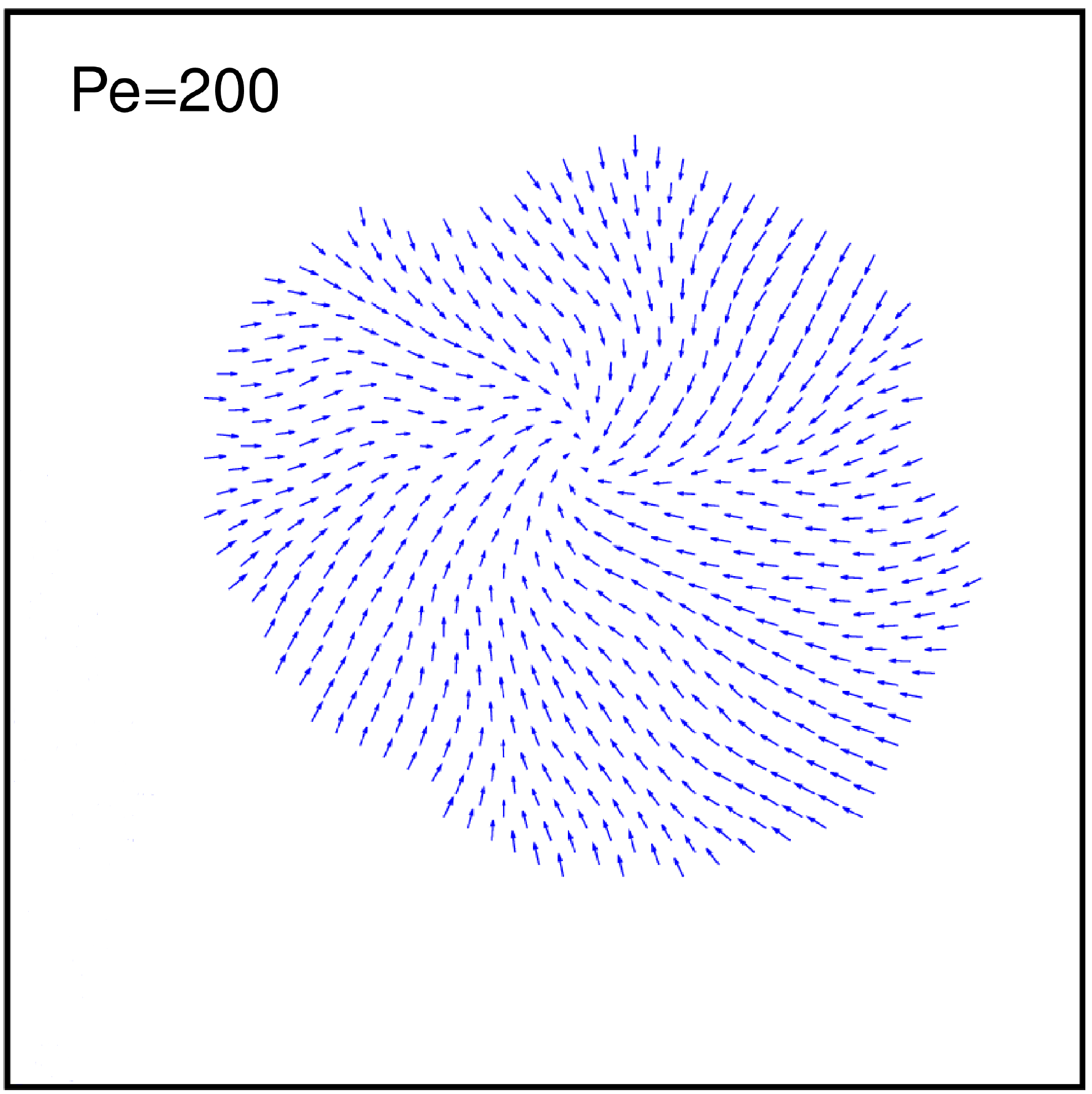}}
	\\
    \end{tabular}
    \vspace{0.2cm}
  \caption{Local polarisation field ${\mathbfcal{P}}$  at Pe = $10,40,100,200$ in the $25\%-75\%$ line.
  The four clusters shown are the ones surrounded by black squares in the lower row in Fig.~\ref{fig:pdfs_pol_25-75}. 
  Increasing Pe, the clusters exhibit a spiralling pattern with an increasing angle between ${\mathbfcal{P}}$ and the vector 
  pointing toward the cluster centres, see also Fig.~\ref{fig:orient_pdf}.
}
\label{fig:topology_polarization}
\end{center}
\end{figure*}

\begin{figure*}[h!]
 \begin{center}
\resizebox{0.8\textwidth}{!}{
  \includegraphics{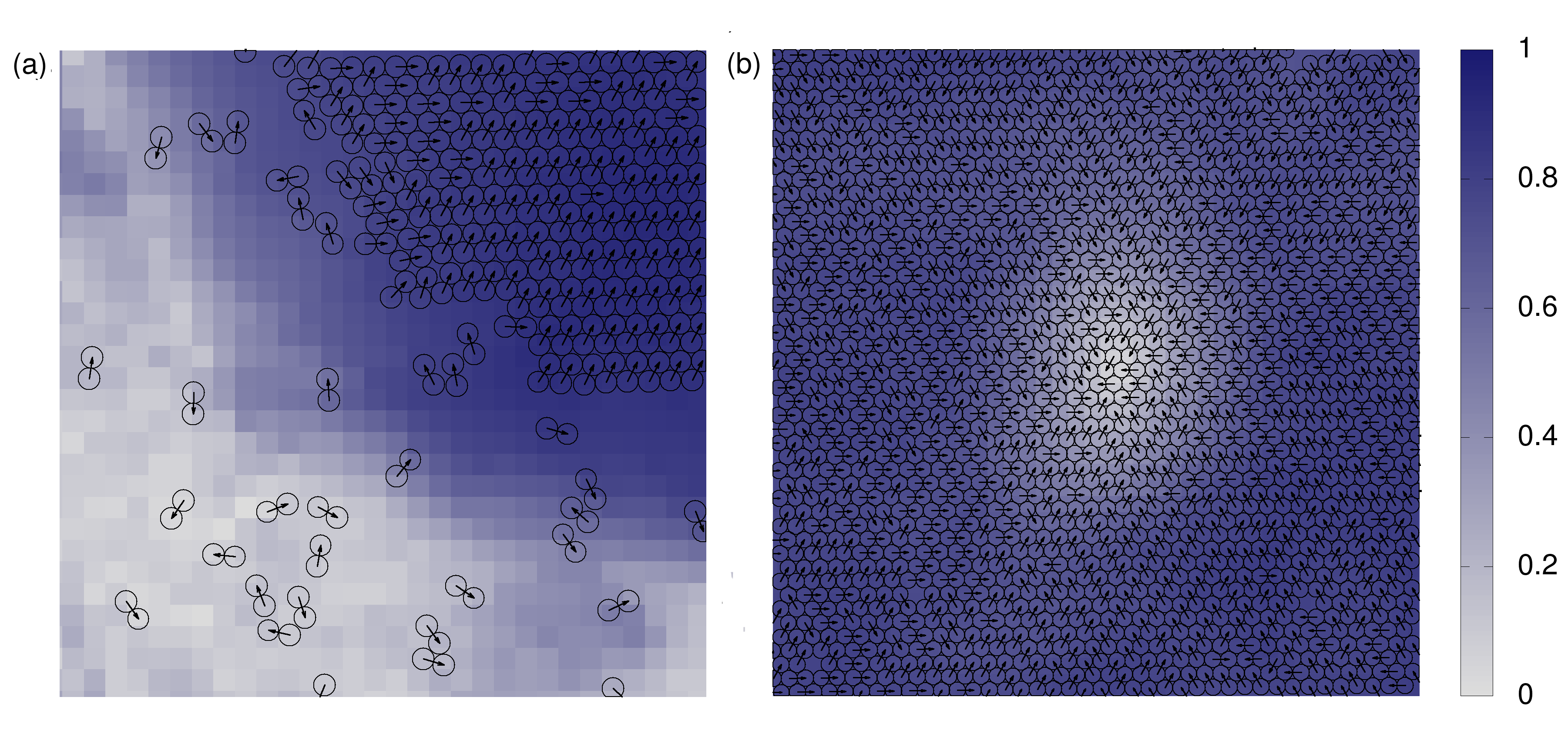}
}
 \end{center}
\caption{Enlargements of the regions surrounded by cyan squares in the snapshot of the system at Pe = $40$, ${\phi=0.310}$ 
(along the $25$\%-$75$\% line), shown in Fig.~\ref{fig:pdfs_pol_25-75}. The colour scale is associated to modulus of the local polarisation field. 
The arrow indicates the tail-to-head direction on each dumbbell. (a) Interface between the cluster and the disordered phase. 
(b) Centre of the cluster.}
\label{fig:cluster_polarization}
\end{figure*}

As the activity increases (Pe $\geq$ 20)
the probability distributions, both on the 50\%-50\% and 25\%-75\% lines, exhibit a two peak structure showing that 
activity tends to induce an effective polar interaction between dumbbells, reflected 
in the emergence of large ordered patterns in the clusters polarisation field.
While along the 50\%-50\% lines the coarse-grained polarisation field remains typically with no polar order (white regions) in the central part of the clusters  and ordered on the border,
along the 25-75\%
line and Pe $\geq$ 40
%, if the activity is strong enough, 
the peak corresponding to random orientation tends to disappear.
In these cases, the disordered region at the centre of each cluster is reduced to a small white spot, 
otherwise stated, a defect for the polarisation field. A deeper look into the local polarisation field, see Fig.~\ref{fig:topology_polarization}, 
reveals the emergence at Pe = 40 of an aster structure with a
 %In these cases, at the centre of each cluster  there is a white spot that represents a region with no polar order,
%otherwise stated, a defect for the polarisation field. 
%At Pe = 40, see panel (b) in Fig.~\ref{fig:topology_polarization}, the polarization field shows an aster structure with a 
charge equal to $q=1$, see panel (b).
This feature can be better appreciated by showing   enlargements of a cluster's snapshot, overlapped with the coarse grained polarisation field,  close to the interface where dumbbells point inward (Fig.~\ref{fig:cluster_polarization} (a)), 
and at the centre where the defect appears (Fig.~\ref{fig:cluster_polarization} (b)).
At larger activity (Pe = 100, 200), the  aster evolves into a spiralling pattern, see panels (c) and (d) in Fig.~\ref{fig:topology_polarization} for two enlarged clusters. Notice that this spiralling pattern is associated to an increased angle between the defect and the local polarisation field, that will be characterised later in this paper, see for instance Fig.~\ref{fig:orient_pdf}. 

%\begin{figure*}[h!]
%\begin{center}

\begin{figure*}
\vspace{0.4cm}
\begin{center}
\resizebox{2\columnwidth}{!}{
  \includegraphics{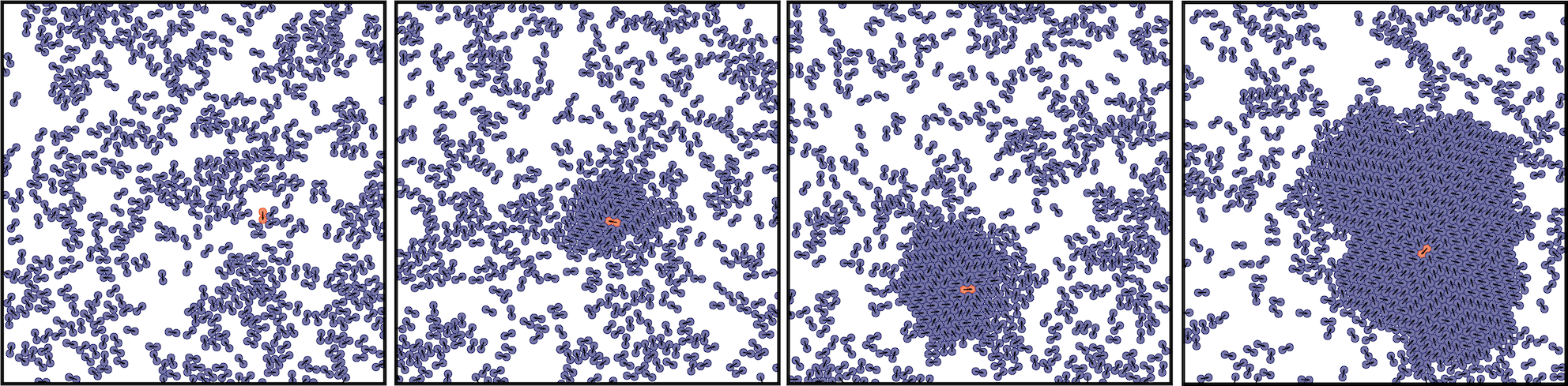}
}
\end{center}
\vspace{0.4cm}
\caption{Snapshots taken from a single run at four different times, $t = 66000 \, \tau, 67000 \, \tau,\ 72000 \, \tau,\ 78000 \, \tau$. 
They show how the cluster is formed around the `central' dumbbell coloured in 
red that triggers aggregation. The parameters are Pe = $40$ and ${\phi=0.310}$.}
\label{fig:cluster_formation}
\end{figure*}

To clarify the reason for the appearance of the aster pattern with  a defect at the cluster's centre 
we can trace back the evolution of one of the dumbbells that is located near the defect, at Pe = 40. 
In this way we can go back to the event that triggered the MIPS in the first place. In Fig.~\ref{fig:cluster_formation} a series of snapshots of the formation 
of the cluster are shown, with the traced dumbbell coloured in orange. 
The cluster formation process starts with multiple head-to-head 
dumbbells  collisions  (thus without polar order). After the initial nucleation, other dumbbells start to 
segregate on the cluster, with a pathway that is necessarily on average directed towards the cluster's centre.  Larger  activities allow the initial colliding
direction to remain fixed after segregation. This  allows the polar field to be ordered in an aster (or spiral) pattern,  and not disordered as observed at the lower value  Pe = 10.

In short, the evolution of the polarisation with Pe is such that
\begin{itemize}
\item[$\bullet$]
there is no polarisation at small Pe, the dumbbells within the cluster orient along the 
radial direction for intermediate Pe, and the clusters show a spiralling structure at larger Pe.
Therefore, in addition to hexatic order,  present for all strength of activities in the MIPS region~\cite{Cugliandolo2017}, we observe in this paper that polar order is also relevant
and differentiates between clusters at small and high activity.
\end{itemize}

\subsubsection{Correlation functions}
 
\begin{figure}[h!]
	\begin{center}
	\resizebox{\columnwidth}{!}{\includegraphics{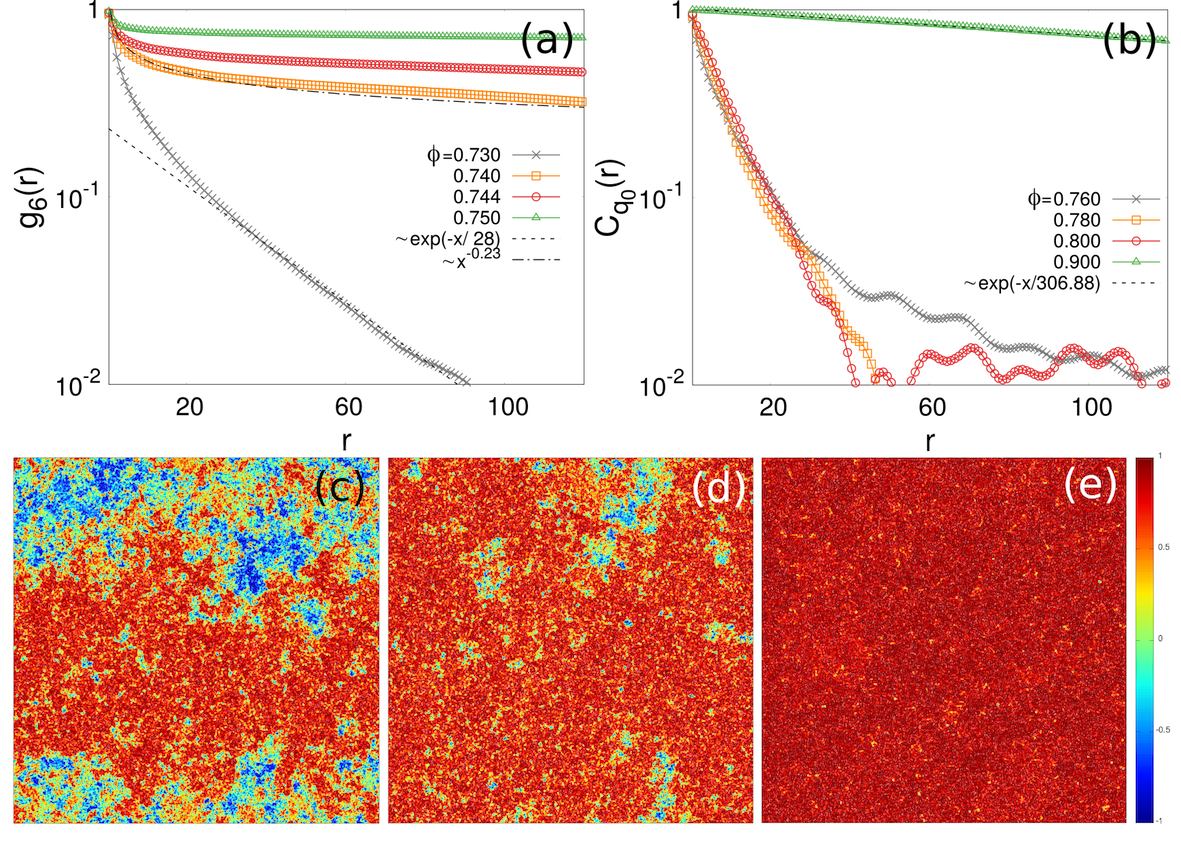}}
	\end{center}
\caption{Orientational correlation functions (a) and positional correlation functions (b), as defined respectively in~Eq.~(\ref{eq:hec_corr_func}) and Eq.~(\ref{eq:position-correlation}) for Pe = 0  and different values of the packing fraction indicated in the keys, in log-lin scale. The highest density with an exponential decay and the lowest density with power-law decay are fitted with a dotted and a dashed line, respectively. For (b) the latter is absent. Above, three snapshots representing local 
hexatic parameter through the methods described in text, for Pe  = 0 and $\phi=0.740$ (c), $0.750$ (d), $0.800$ (e).
}
\label{fig:correlation_functions_Pe0}
\end{figure}

\begin{figure}
\vspace{0.4cm}
	\begin{center}
	\resizebox{\columnwidth}{!}{\includegraphics{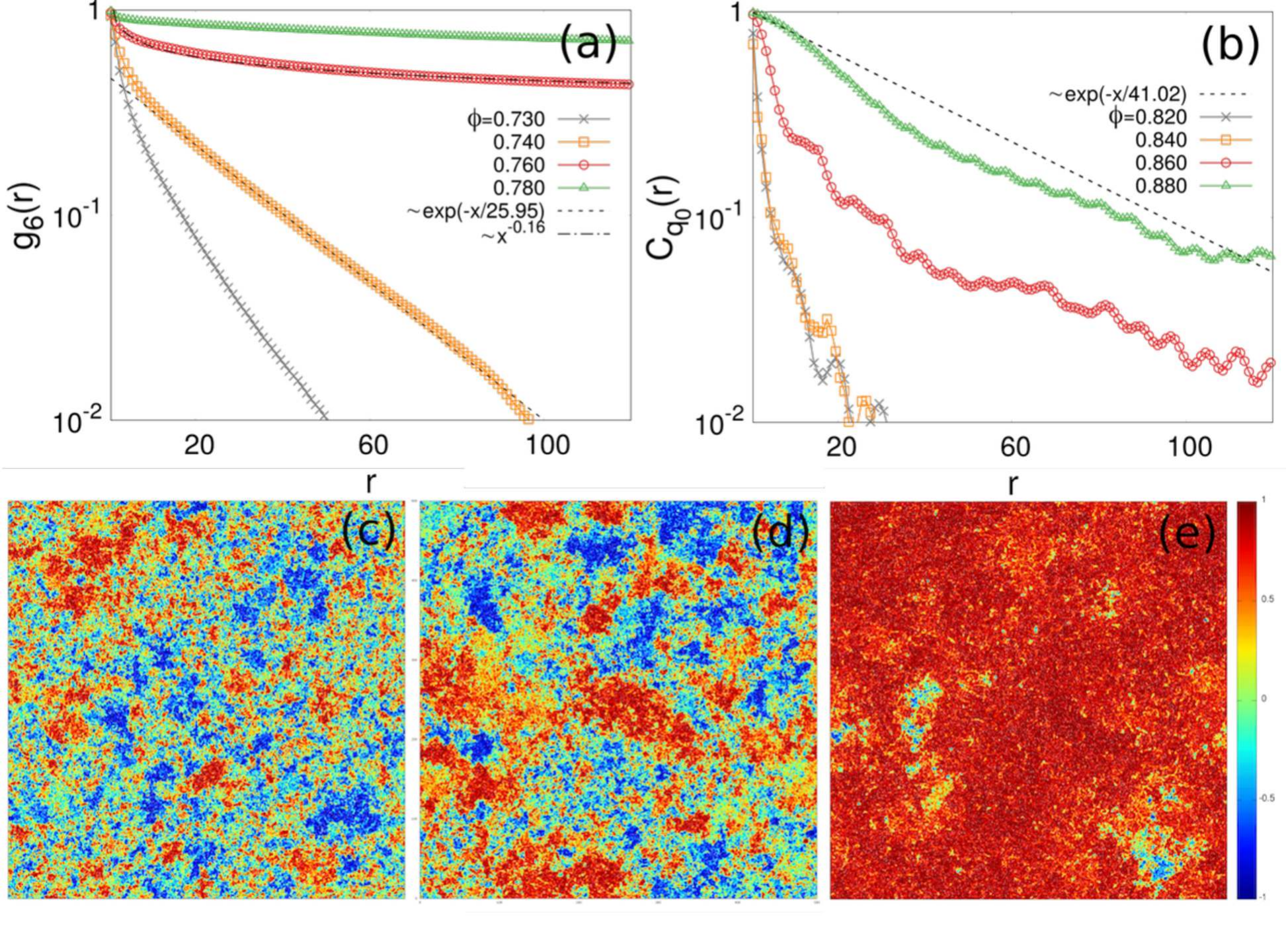}}
	\end{center}
\caption{Orientational correlation functions (a) and positional correlation functions (b) for Pe = 2, in log-lin scale and with exponential and power-law fits provided. 
Below, typical snapshots for Pe = 2  and $\phi=0.730$ (c), $0.740$ (d), $0.800$ (e).}
\label{fig:correlation_functions_Pe2}
\end{figure}

\begin{figure}
	\begin{center}
	\resizebox{\columnwidth}{!}{\includegraphics{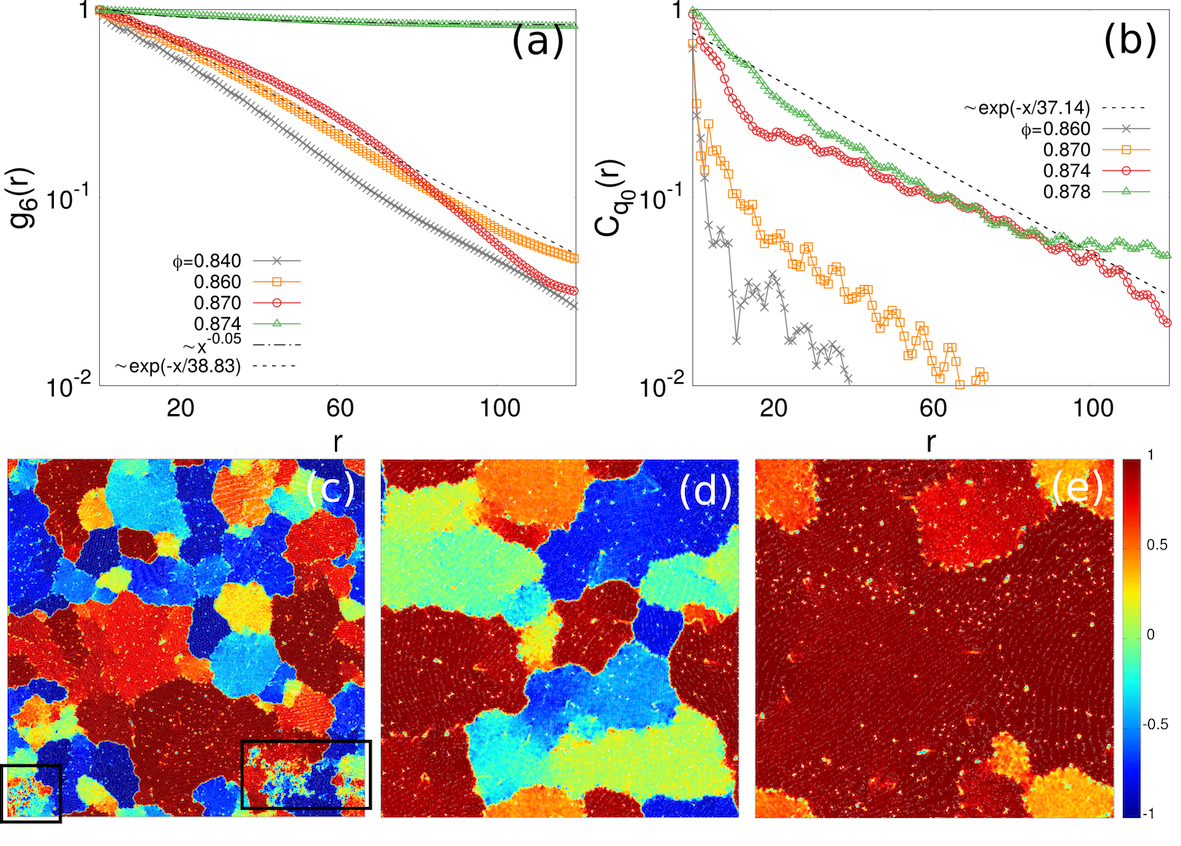}}
	\end{center}
\caption{Orientational correlation functions (a) and positional correlation functions (b) for Pe = 10, in log-lin scale and with exponential and power-law fits provided. 
Below, snapshots for Pe = 10  and $\phi=0.860$ (c), $0.872$ (d), $0.874$ (e). In the snapshot (c) the two disordered regions are highlighted by the two black squares.}
\label{fig:correlation_functions_Pe10}
\end{figure}

The orientational and positional correlation functions are shown in Figs.~\ref{fig:correlation_functions_Pe0},~\ref{fig:correlation_functions_Pe2} and~\ref{fig:correlation_functions_Pe10}, side-by-side, for three activities Pe = 0, 2, 10, and various packing fractions indicated in the keys.
Below these plots, typical configurations in the stationary regime of various representative pairs (Pe, $\phi$) are displayed, in order to better understand the behaviour of the correlations.
All  correlation functions are plotted on a log-lin scale, in order to clearly identify transitions between exponential and power-law decays. Moreover, the highest density with exponential decay and the lowest density with power-law decay (if present) are fitted with a dotted and a dashed lines, respectively.

For all the values of the activity, Pe = 0, 2, 10, the orientational correlation function  changes from an exponential decay at low $\phi$ to an 
algebraic one at high $\phi$. We interpret this change as being due to the fact that the density parameter is moving across the coexistence region, 
so that the region above coexistence is characterised by quasi-long-ranged orientational order (see Sec.~\ref{subsec:2d-melting}). 
This same behaviour was found in the first order liquid-hexatic transition of passive spherical particles~\cite{BeKr11} and allows us to claim that, 
\begin{itemize}
\item[$\bullet$]
both in equilibrium and for all positive values of the activity, the orientational correlations 
suggest that there is coexistence between a disordered phase and a hexatically ordered one, depicted with the grey gradient in the phase diagram of Fig.~\ref{fig:phase_diagram}.
\end{itemize}
On the other hand, positional correlation functions, not shown in~\cite{Cugliandolo2017}, never exhibit  algebraic decay, except in the close-packing limit (see, for example, the curve at Pe = 0 and $\phi=0.900$ in Fig.~\ref{fig:correlation_functions_Pe0} (b), which can be still fitted by an exponential). 
\begin{itemize}
\item[$\bullet$]
Since monomers are constrained to be attached in pairs, they cannot arrange on a triangular lattice at any $\phi<\phi_{\rm cp}$, forcing the positional correlations of the monomers to decay exponentially.
\end{itemize}
See for example in Fig.~\ref{fig:correlation_functions_Pe0} (b) than even at $\phi=0.800$ the positional order is short-ranged, although the system is already orientationally ordered as shown by the snapshot in Fig.~\ref{fig:correlation_functions_Pe0} (e).
 With positive activity we can also observe some mixed cases, {\it e.g.} the curve at Pe = 2, $\phi=0.880$ in Fig.~\ref{fig:correlation_functions_Pe2} (b), with an initial algebraic decay and an exponential tail, since activity is able to drive the system to form very close-packed ordered domains together with disordered regions among them 
 (see the snapshot in Fig.~\ref{fig:correlation_functions_Pe2} (e)). Increasing activity further, the upper limit of the coexistence is pushed towards higher values of the
 packing fraction. This makes the analysis of the phase above coexistence much more demanding, because of the slowing down of the relaxation time. Snapshots in Fig.~\ref{fig:correlation_functions_Pe10} (c) and (d) are one below (note the small disordered hole at $\phi=0.860$) and one above the coexistence 
 limit and they are both in a metastable polycrystalline state explaining why both the orientational order and the positional one decay within the system size.
The exact nature of the transition between the hexatic and solid phases for the dumbbell system and its location in the phase diagram are still  open questions. We aim to go deeper into them in our future works.
 
\section{Dynamic  properties}
\label{sec:dynamic}

For the moment we have focused on the analysis of the structural properties of the system, {\it via} the analysis of the 
local density, local hexatic order parameter, their correlation functions, {\it etc.} but we have not studied the 
behaviour of the velocity field. In this Section we investigate various aspects  of the dumbbell velocities
and the translational and rotational MSD in the finite density system. The dynamic observables are also  expected to  
strongly depend on where they are evaluated in the phase diagram. Once again, we choose
parameter values on curves of constant proportion of the dense and loose phases in the region of 
coexistence, the most interesting region of the phase diagram.

\subsection{The kinetic energy} 

\begin{figure}[h!]
\resizebox{\columnwidth}{!}{
  \includegraphics{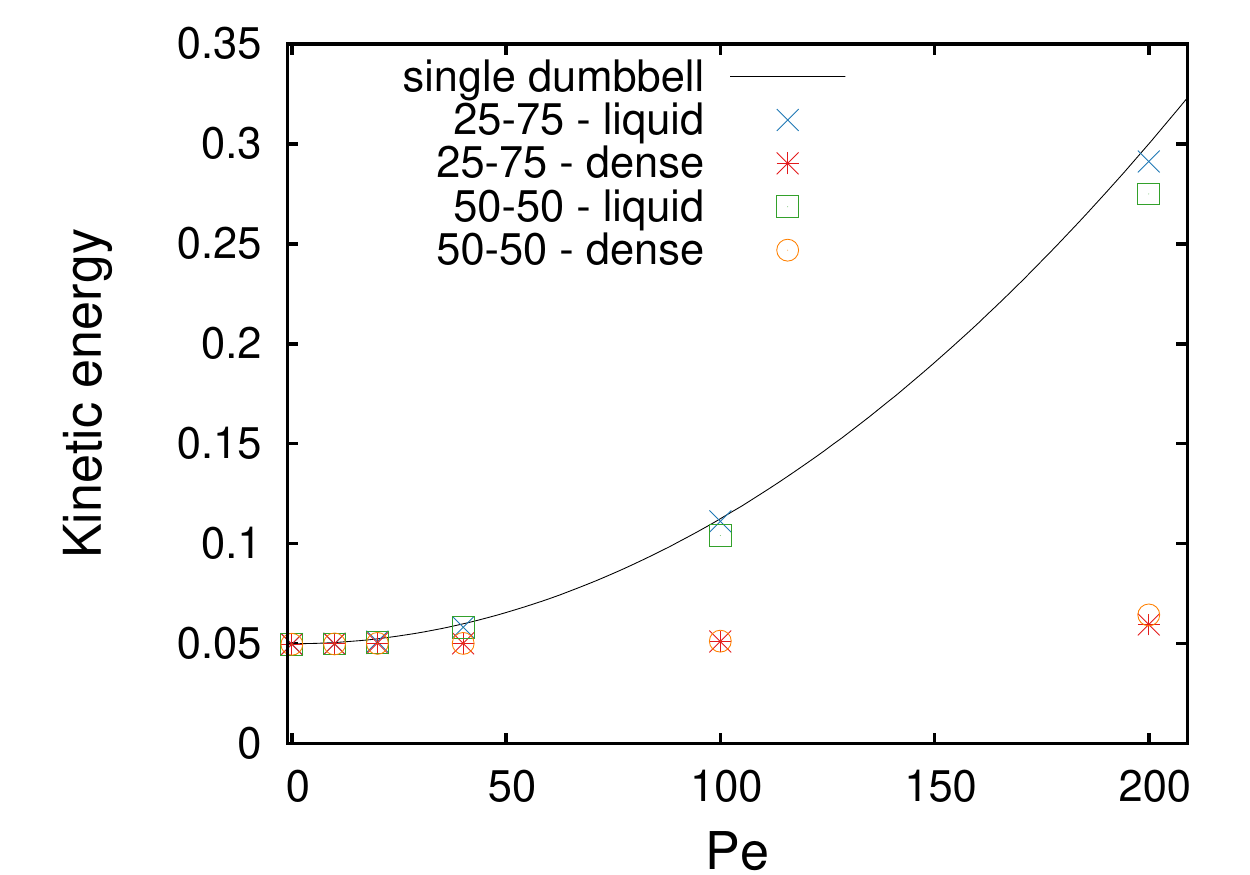}
}
\caption{The averaged kinetic energy of particles in the dense and dilute phases, for parameters on the $25$\%-$75$\% and $50$\%-$50$\%
curves. For comparison, the single dumbbell kinetic energy, for the same parameters, is shown with a black line, see Eq.~\ref{eq:Ekin}.
%Note that the global packing fractions are, approximately, $\phi = 0.25$ on the curve with 25\% dense phase and $\phi = 0.5$ on the curve with 50\% dense 
%phase, for Pe = 200.
}
\label{fig:Ekin}
\end{figure}

In this work, we measured   the kinetic energies of the particles that belong to the dilute phase separately from the one of those 
that belong to the dense
phase, along the $50$\%-$50$\% and the $25$\%-$75$\% curves in the phase diagram. The data are shown in 
Fig.~\ref{fig:Ekin}. The kinetic energy of the beads that form dumbbells in the liquid phase grows as a monotonic function of 
Pe in agreement with  the expected behaviour for a single dumbbell,
the averaged kinetic energy of which is  shown with a solid black line in the same figure
(a Pe$^2$ dependence, see Eq.~(\ref{eq:Ekin})). Indeed,
\begin{itemize}
\item[$\bullet$]
the dumbbells in the dilute phase are basically free, since their kinetic energy 
is very close to the one of independent dumbbells all along the Pe values.
\end{itemize}
The kinetic energy of the particles in the dense phase behaves very differently
and, on the scale of the figure, their $E_{\rm kin}$ barely increases from 0.05 to 0.06.
\begin{itemize}
\item[$\bullet$]
The kinetic energy of the particles in the dense phase has a very weak growth, 
due to the fact that the mobility is suppressed inside the clusters and that clusters are massive and move very slowly.
\end{itemize}
A distinction between the kinetic energy of the two kinds of dumbbells 
becomes relevant at around Pe = 30, in the scale of the figure.

\subsection{The velocity field}

\begin{figure}[h!]
(a) \hspace{2.25cm} (b) \hspace{2.25cm} (c)
\begin{center}
\resizebox{\columnwidth}{!}{\includegraphics{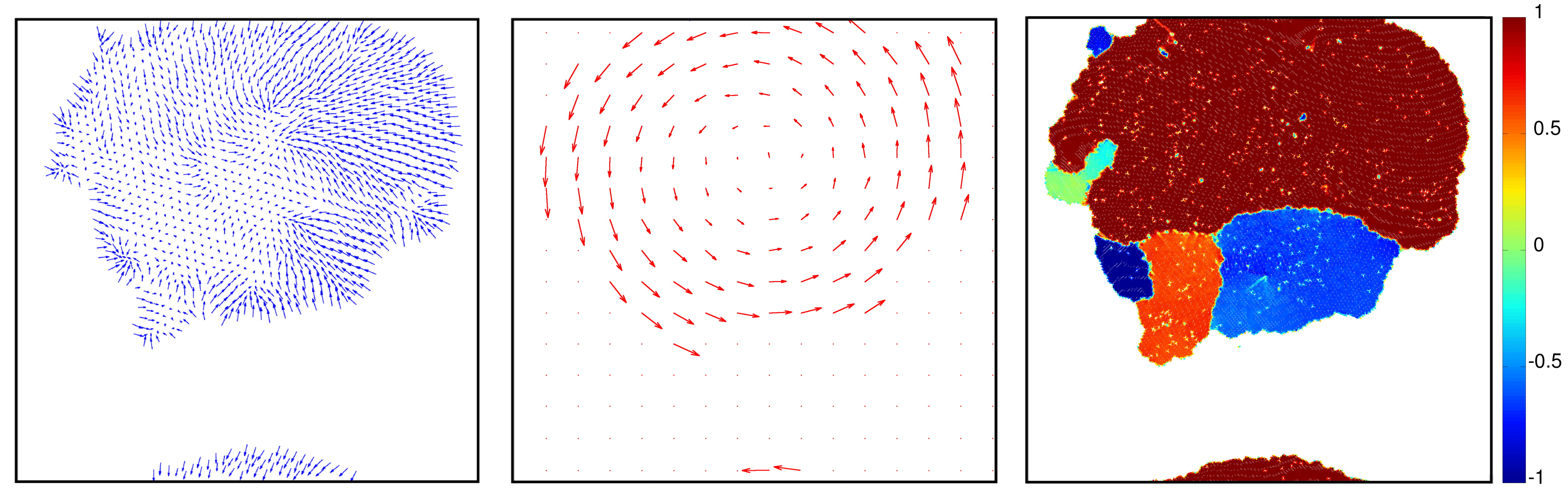}
}
\end{center}
\caption{The polarisation map 
(a), the  instantaneous map of the coarse-grained velocity field (using $R= 20\, \sigma_{\rm d}$) (b);
and the local hexatic parameter (c) of the active system with  Pe = 200 on the 50\%-50\% line. 
}
\label{fig:velocity_field}
\end{figure}

We evaluated the velocity field by first averaging over a typical time 
($\tau_{LJ}=1$) the instantaneous velocity of each  bead relative to the centre of mass of the full 
system and then coarse-grained these values  over square
plaquettes with side $R$. We used different values of $R$ depending on the typical cluster size. 
For Pe = 40, 100, 200  the choice ${R=20 \, \sigma_d}$ seems appropriate, 
while for smaller Pe we had to use smaller values of $R$ to see clearly the velocity field. 
Figure~\ref{fig:velocity_field}
shows the polarisation map (in panel (a)), a snapshot of the coarse-grained velocity field  (in panel (b)) 
together with the local hexatic parameter of the same configuration (in panel (c)) of a selected part of a 
system with Pe = 200 on the 50\%-50\% curve. The system is clearly 
phase separated between a single cluster made of seven regions with different hexatic order and 
the rest of the area with very few free dumbbells making this region almost empty. It is also clear from the 
velocity field that 
\begin{itemize}
\item[$\bullet$]
the cluster at Pe = 200 is turning around its centre and there is no signature of the 
domains with different hexatic orders, which compose the cluster, in the velocity field. The rotation of the clusters is a general feature at sufficiently
high Pe and this suggests a more detailed study that will be done in the following, with the help of the analysis of the
enstrophy.
\end{itemize}

\subsection{The enstrophy}

\begin{figure*}[!htb]
\begin{center}
\resizebox{1\textwidth}{!}{
\includegraphics{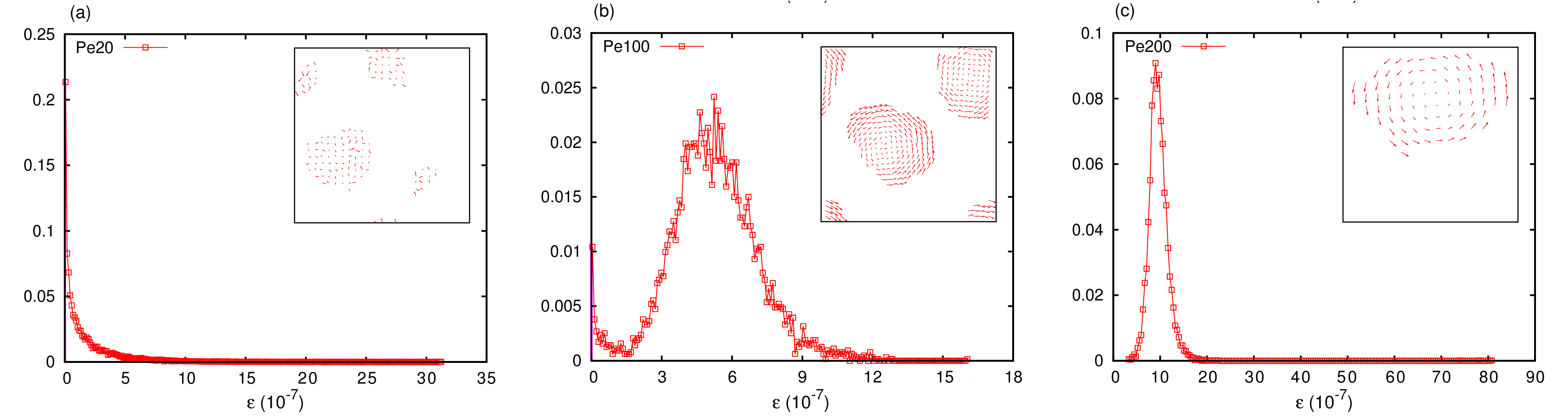}
}
\end{center}
\caption{Main plots: the enstrophy pdfs at Pe = $20$,  $100, \ 200$
along the 50\%-50\% curve. 
Inserts: coarse-grained velocity field in four representative instantaneous configurations  at the same parameters. 
The peaks appearing in the pdfs at Pe = 100, 200, are associated to clusters undertaking a rotational motion.
}
\label{fig:pdfs_enstrophy_50-50}
\end{figure*}

\begin{figure*}[!htb]
\begin{center}
\resizebox{1\textwidth}{!}{
\includegraphics{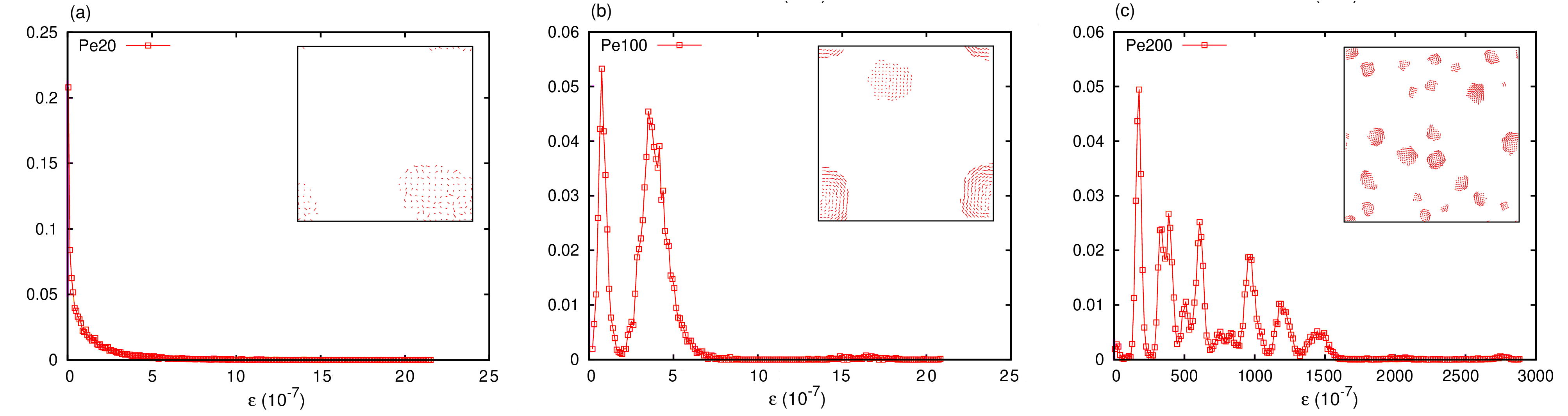}
}
\end{center}
\caption{Main plots: the enstrophy pdfs at Pe = $20$,  $100, \ 200$
along the 25\%-75\% curve. 
Inserts: coarse-grained velocity field in four representative instantaneous configurations  at the same parameters. 
The peaks appearing in the pdfs at Pe = 100, 200, are associated to clusters undertaking a rotational motion.
}
\label{fig:pdfs_enstrophy_25-75}
\end{figure*}

The enstrophy is defined as
 \begin{equation}
\epsilon=\frac{1}{2}\sum_{\bold{r}}|\boldsymbol{\omega}(\bold{r})|^2
\; ,
 \end{equation}
 where ${\bold{r}}$ locates the grid points, defined in the previous section, and ${\boldsymbol{\omega}={\boldsymbol{\nabla}}\times\bold{v}}$ is the vorticity vector, with 
 $\bold{v}$ the coarse-grained velocity field, calculated as described in the previous section.
While the kinetic energy gives a measure of the strength of flow in the system, the enstrophy is a 
measure of the presence of vortices in the velocity field 
and it can be used to understand whether clusters rotate driven by activity.
In order to measure meaningful values of the enstrophy, the correct coarse-grained scale of the velocity field must be carefully chosen. In particular, it must be larger than the particle’s size and smaller than the typical size of the aggregated phase.  
%We should be able to associate a specific enstrophy value for each rotating cluster and distinguish it from the background thermal noise. 
We found appropriate lengths between 10 and 20$\sigma_d$ (we opted for 10$\sigma_d$).

In Figs.~\ref{fig:pdfs_enstrophy_50-50} and \ref{fig:pdfs_enstrophy_25-75} we show the pdfs of the enstrophy, 
evaluated over several independent and  stationary configurations. The  panels display the results  for systems at different 
Pe numbers, Pe  = 20,  100, 200,  along the 50\%-50\% and the 25\%-75\% curves in the phase diagram. 
The inserts are maps of the velocity field and 
show its typical behaviour. 
The probability of finding a non-zero value of the enstrophy decreases smoothly for low Pe number with an exponential tail (not distinguishable in the plot), with the snapshots  of the velocity field at 
Pe = 20, 40 (not shown) suggesting  a lack of coherent rotation. 
It can be checked that the exponential decay at Pe = 20, 40 is very close  to the one at Pe = 0, meaning the enstrophy values are purely associated to the thermal noise.
Since we do not expect any relevant vortical behaviour in the passive case,  the noise level can be approximately estimated  by measuring the enstrophy distribution at Pe = 0. 
 Moreover, as noticed in Sec.~\ref{subsec:polarisation_results}, for low values of the Pe number the coarse-grained polarisation field looks completely disordered, 
while at Pe = $40$ the dumbbells tend to point along the radial direction towards the centre of the cluster without exhibiting a spiralling pattern. In both 
cases these features seem to prevent a coherent rotation. At higher Pe (see the cases at Pe = 100, 200), 
the probability distribution function of the enstrophy develops a multi-peak structure
depending  on the number and size of the clusters on scales  clearly distinguishable from the noise contributions. One can say that 
\begin{itemize}
\item[$\bullet$]
For high values of the activity (Pe $\ge 100$) the dumbbells arrange in spirals, the clusters in the system undertake a rotational 
motion and the probability distribution function of the enstrophy develops a multi-peak structure (without an exponential tail,  typical of lower Pe) associated to the  rotating clusters
\end{itemize}
For intermediate values of the activity such as Pe = $70$, we can also 
observe cases of a coherent, even if less evident, rotation.

In Ref.~\cite{Suma14} the authors presented a minimal mean-field framework within which it is possible to understand the 
physics of these rotating aggregates. It was noticed in this Letter that, in contrast with the case of motility-induced clusters observed with 
spherical particles, steric interactions quench the polarisation of the dumbbells in the rotating cluster. Each particle exerts a 
local torque which is balanced by the drag on the cluster, sustaining rotations. By comparing the 
velocity field and the polarisation we found here a substantial distinction: 
\begin{itemize}
\item[$\bullet$]
the velocity field exhibits a vortex pattern while the polarisation one is a spiral.
\end{itemize}
In the same context, it was  inferred 
that the angular velocity scales with cluster size as ${1/R}$ and  found that its order of magnitude is around ${10^{-4}}$. If one accepts the 
simplifying assumption that the clusters perform a uniform motion, the enstrophy is the angular velocity squared and, in this approximation, 
its order of magnitude is compatible with the data shown in Figs.~\ref{fig:pdfs_enstrophy_50-50} and \ref{fig:pdfs_enstrophy_25-75}.

In order to measure the mean angular velocity of the rotating clusters and the orientations of the dumbbells inside each cluster, it is necessary to distinguish between the particles belonging to a cluster, characterised by high-density, and the free particles moving in low density regions. 
To perform this task, we used a density-based clustering algorithm, DBSCAN (Density-Based Spatial Clustering of Application with Noise)~\cite{Ester},
%we first separate the aggregates from the background using the algorithm
that gathers points closely packed together and that marks as noise points lying 
in low-density regions. We associate a particle to a cluster when it has at least $10$ nearby particles within a circle of radius $2\sigma_d$ around its centre, and in this way we are able to locate all the  cluster aggregates (and the particles contained in each of them).

\begin{figure}[h!]
\resizebox{\columnwidth}{!}{
  \includegraphics{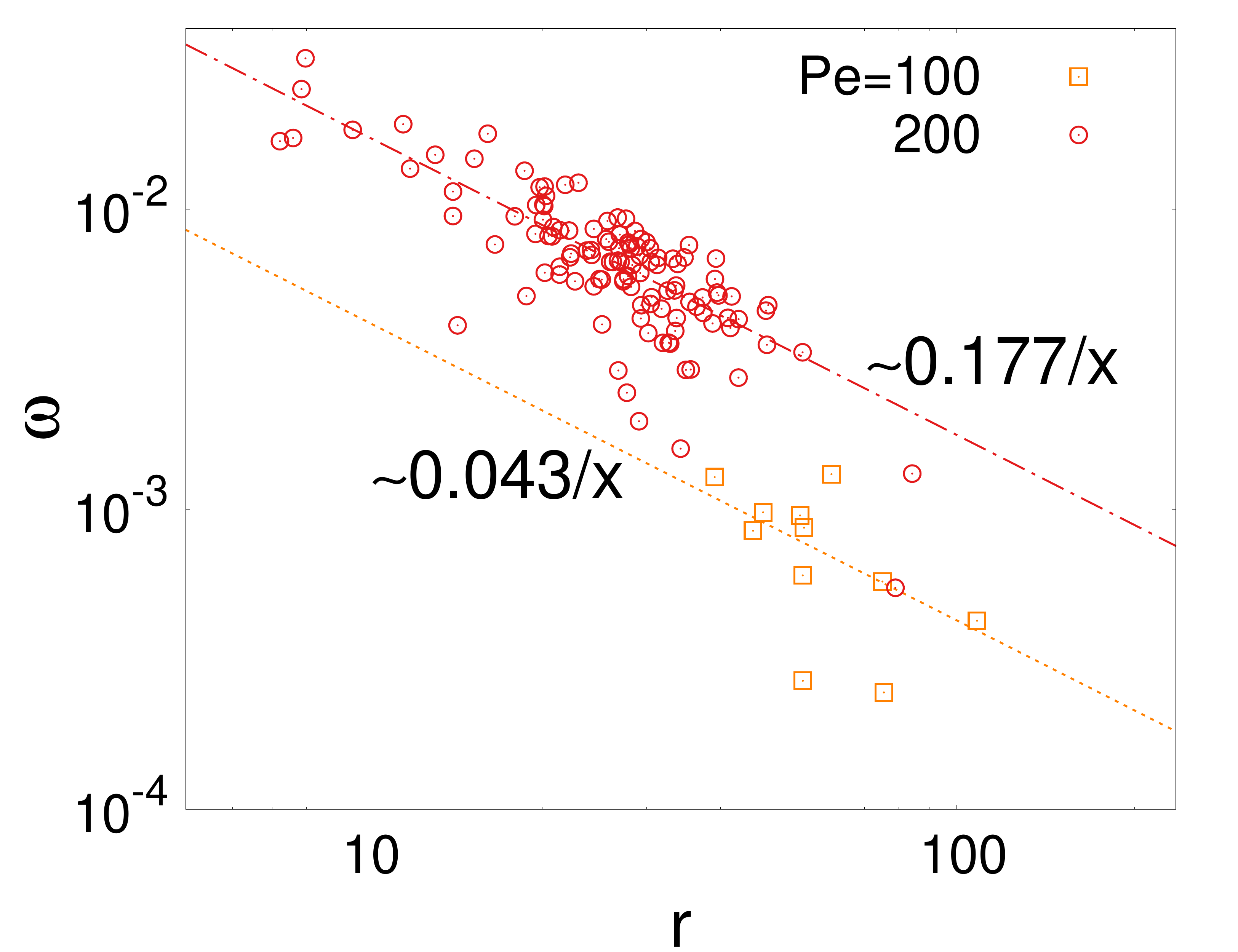}
}
\caption{Log-log scatter plot of the angular velocity $\omega$ of the clusters versus their radius $r$, at 
Pe = $100,\ 200$ along the 25\%-75\% curve. A fit of the  function $a/r$ for each Pe is shown as an inclined line, 
with $a=0.043, 0.177$ for Pe = $100,\ 200$, respectively.
}
\label{fig:omega_vs_r}
\end{figure}

Once the clusters are identified, we first associate a radius to them, by calculating the radius of a disk with the same moment of inertia.
Then we calculate
the angular moment with respect to the cluster's centre of mass. In this way, we easily get an estimate of the angular speed of the 
rotating clusters around their centre of mass. In Fig.~\ref{fig:omega_vs_r} we show the results  obtained at Pe = $100,\ 200$ along the 25\%-75\% curve. 
\begin{itemize}
\item[$\bullet$]
The angular velocity scales with cluster size as ${1/R}$,
\end{itemize}
\noindent
as found in~\cite{Suma14}, but we underline that in the case presented here we considered aggregates that spontaneously formed during the motility-induced phase separation, and are not instead created artificially as in~\cite{Suma14}.

\begin{figure}[h!]
\resizebox{\columnwidth}{!}{
  \includegraphics{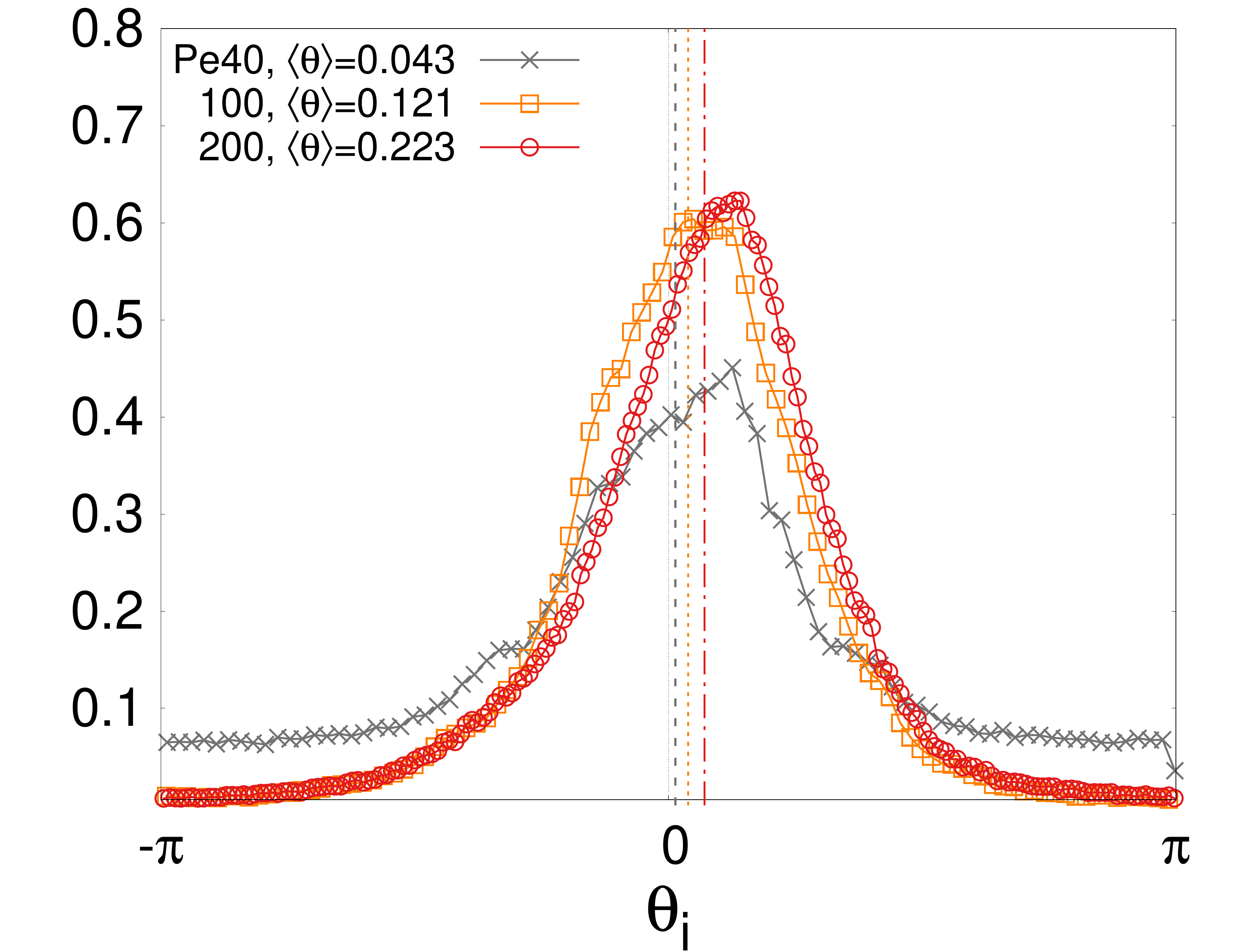}
}
\caption{Pdf of the orientation of the dumbbells in 
clusters, in systems 
at Pe = $40,\ 100, \ 200$ along the 25\%-75\% curve. The orientation is defined by the angle 
between the head-to-tail vector of the dumbbells and the vector pointing from the centre of mass of the dumbbell to the centre of the 
cluster.
}
\label{fig:orient_pdf}
\end{figure}

If we define the orientation of each dumbbell inside a cluster as the angle between its head-to-tail vector and  the vector pointing from the centre 
of mass of the dumbbell to the centre of the cluster, we can give a quantitative measure of the different patterns exhibited by the polarisation field 
in Fig.~\ref{fig:topology_polarization} and also understand how, at high Pe values, a net torque can favour the rotation of clusters. 
In Fig.~\ref{fig:orient_pdf} the pdfs of the dumbbells orientations are shown for  
systems at Pe = $40,\ 100,\ 200$. 
The distributions are  asymmetric and present a non zero average angle which grows with Pe, matching what observed in the polarisation field, see Fig.~\ref{fig:topology_polarization}. 

Therefore, our numerical results for high values of the active force are in perfect agreement with the 
prediction in~\cite{Suma14}, where
the simulations were initialised with a high-density cluster formed placing the dumbbells in spirals at given orientations.
At low Pe numbers the dynamics looks different. When the dumbbells point along the radius of the cluster, towards its centre, or when 
their directions seem disordered and the polarisation field does not show a spiralling pattern,
no rotation is observed for the aggregates.

\subsection{Mean-square displacements}

The positional mean-square displacement of the centre of mass of the single dumbbell, Eq.~(\ref{eq:MSD-CM-single}),  generalises to 
\begin{equation}
\Delta_{\rm cm}^2(\Delta t) = \frac1{N} \sum_{i=1}^N \langle ({\bold r}_{{\rm cm}_i}(t+\Delta t) - {\bold r}_{{\rm cm}_i}(t) )^2 \rangle
\end{equation}
and the rotational  one, Eq.~(\ref{eq:MSD-theta-single}), to 
\begin{equation}
\Delta_\theta^2(\Delta t) = \frac1{N} \sum_{i=1}^N \langle (\theta_i(t+\Delta t) - \theta_i(t) )^2 \rangle
\end{equation}
in a system with $N$ dumbbells. As we will discuss below, these global measurements can lead to confusing or 
even misleading results in heterogeneous cases with phase co-existence. When this arises, a distinction between particles 
in the dense and liquid phases may be necessary to understand the behaviour of these quantities and, more generally, of the 
system as a whole. This can be done, for 
example, by following the probability distribution functions (pdf) of the displacement along a chosen direction, 
say $x$, at a certain time-lag $\Delta t$:
\begin{eqnarray}
p(\Delta x_{\rm cm}, \Delta t) &=& \frac1{N} \sum_{i=1}^N \delta(\Delta x_{\rm cm}-\Delta x_{{\rm cm}_i}(\Delta t) )
\; , 
\\
p(\Delta_\theta,\Delta t) &=& \frac1{N} \sum_{i=1}^N \delta(\Delta_\theta-{\Delta_\theta}_i(\Delta t))
\; ,
\end{eqnarray}
for a single run, and their average over many repetitions of the dynamics under the same conditions
that will naturally lead to smoother functions.

The  finite density effects on the dynamics, for relatively  low Pe values, that is to say, mostly in the liquid phase, 
were evaluated in~\cite{Suma14,Suma14b}. 
The four dynamic regimes in the centre of mass MSD of a single dumbbell, described in Sec.~\ref{sec:single-dumbbell}, 
are preserved for sufficiently low densities. The translational diffusion constant in the ultimate diffusive regime, $D_A$, 
and rotational diffusion constant, $D_R$, once normalised by the thermal energy scale, $k_BT$, depend only on $\phi$ and Pe. $D_A$ diminishes with
increasing density while, for sufficiently high Pe, $D_R$ first increases and next decreases with $\phi$. Both are enhanced by Pe and 
the functional dependence on this parameter is more involved than just quadratic. However,

\begin{figure}[h!]
\resizebox{\columnwidth}{!}{
  \includegraphics{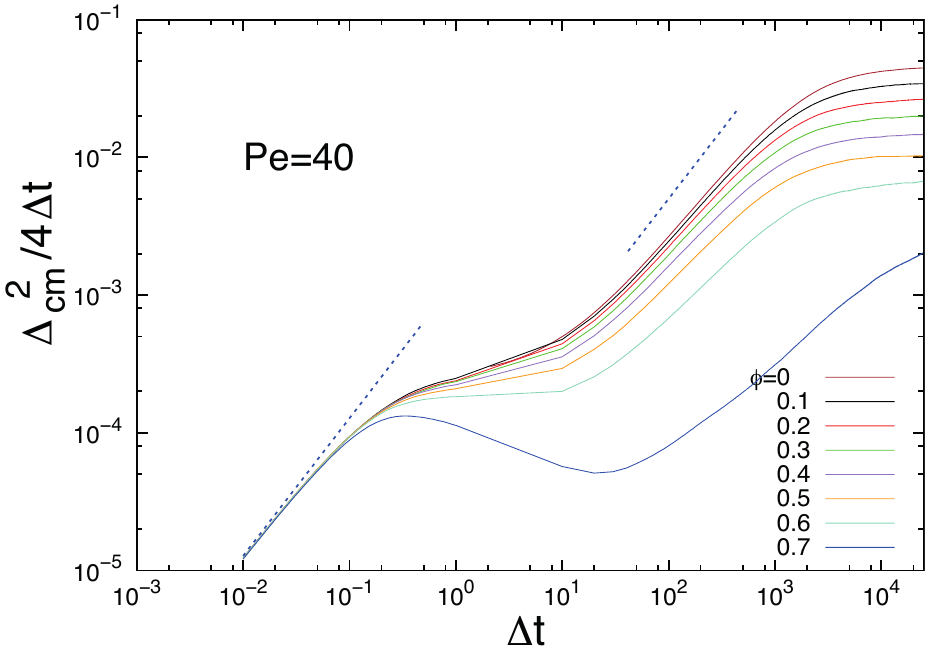}
  }
\caption{The overall dumbbell centre of mass mean-square displacement 
$\Delta^2_{\rm cm}$ as a function of $\Delta t$, for various densities given in the key. 
See the text for a discussion. 
}
\label{fig:MSD-CM-phase-coexistence}
\end{figure}

\begin{figure}[h!]
\resizebox{\columnwidth}{!}{
  \includegraphics{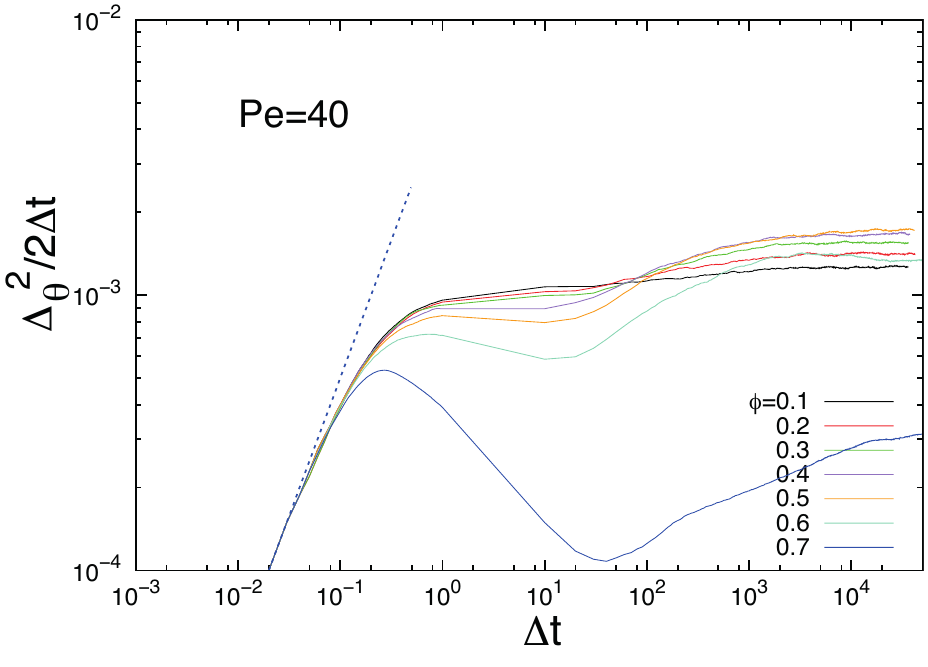}
  }
\caption{The overall dumbbell angular mean-square displacement 
$\Delta^2_\theta$ as a function of $\Delta t$, for various densities given in the key. 
See the text for a discussion.
}
\label{fig:MSD-angle-phase-coexistence}
\end{figure}

\begin{itemize}
\item[$\bullet$]
at sufficiently large packing fraction the density effects are stronger and a qualitative change in the  properties of the 
MSDs are found, as can be seen in Figs.~\ref{fig:MSD-CM-phase-coexistence} and~\ref{fig:MSD-angle-phase-coexistence}
\end{itemize}

The liquid like behaviour, with the four regimes recalled in Sec.~\ref{sec:single-dumbbell}, is progressively 
modified by the increasing density. For the globally denser system, $\phi=0.7$, an anomaly in the first
diffusive and second ballistic regimes in the centre of mass MSD is clear, see Fig.~\ref{fig:MSD-CM-phase-coexistence}.  
The rotational properties are modified even more, with the rotational MSD being strongly affected already at $\phi\simeq 0.35$ and the 
rotational diffusion constant $D_R$ being enhanced by density beyond this value of $\phi$.

\begin{figure}[h!]
\resizebox{\columnwidth}{!}{
  \includegraphics{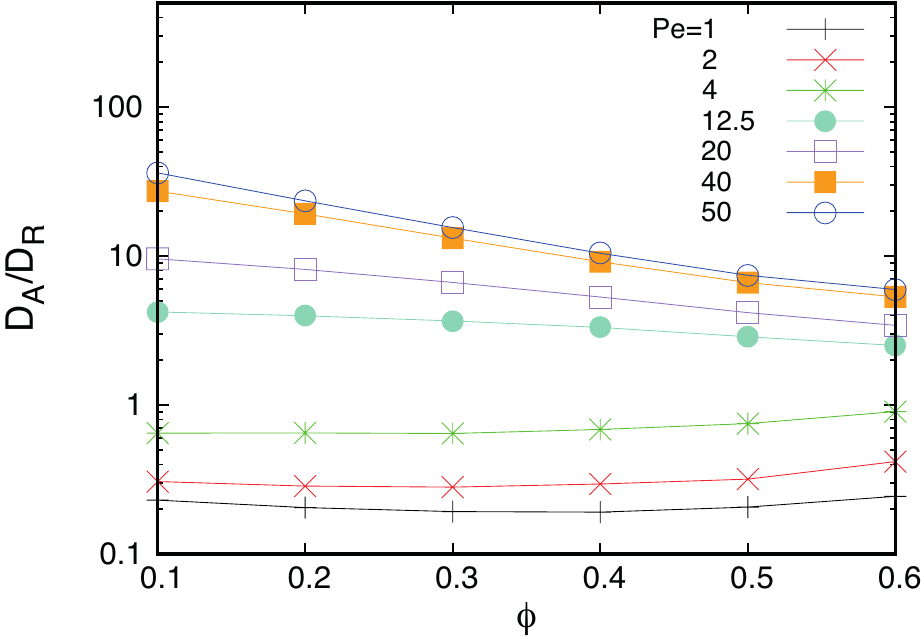}
  }
\caption{The ratio between the translational and the rotational diffusion constants as a function of packing fraction for 
various values of the P\'eclet number given in the key.
}
\label{fig:DA-DR}
\end{figure}

Another view on the global motion of the system is given by the comparison between the centre of mass diffusion constant and the 
rotational diffusion constant, both measured in the last time-delay regime. This is shown in Fig.~\ref{fig:DA-DR} in the form of the 
$\phi$ dependence of the ratio $D_A/D_R$ for different Pe numbers, ranging from Pe = 1 to Pe = 50. (We excluded from this figure 
the case $\phi=0.7$ for which the last diffusive regime does not establish in the time window of the simulation.) 
The maximal 
value is obtained at small $\phi$ and large Pe and it is around 30. At large Pe, say Pe $ \leq 50$ the ratio weakly decreases with $\phi$ while at small
Pe the trend is the opposite.  Observe that, if we consider the typical values of translational and rotational diffusion coefficients for non-tumbling 
bacterial colonies ~\cite{Elgeti15,valeriani2011colloids}, the experimental ratio $D_A/D_R$ appears slightly higher than our results.

\begin{figure}
\resizebox{\columnwidth}{!}{
  \includegraphics{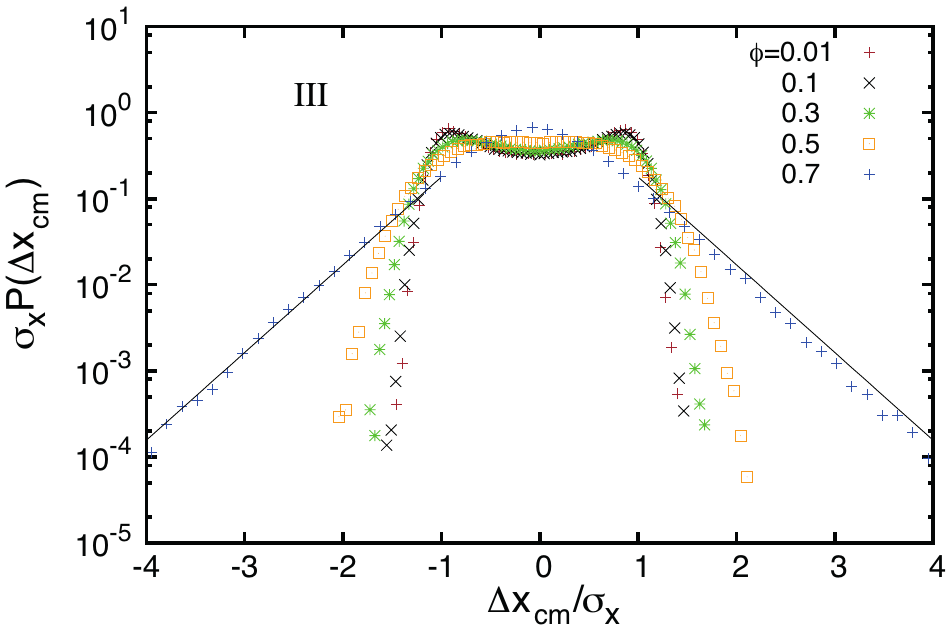}
  }
\caption{The pdf of the centre of mass displacement along the horizontal axis 
at a fixed $\Delta t=t_a/2$ for which the total centre of mass displacement is non-monotonic, 
see Fig.~\ref{fig:MSD-CM-phase-coexistence}. Pe = 40 and densities 
given in the key. $\sigma_x$ is the dispersion of this pdf.
}
\label{fig:pdf-CM-phase-coexistence}
\end{figure}

\begin{figure}
\resizebox{\columnwidth}{!}{
  \includegraphics{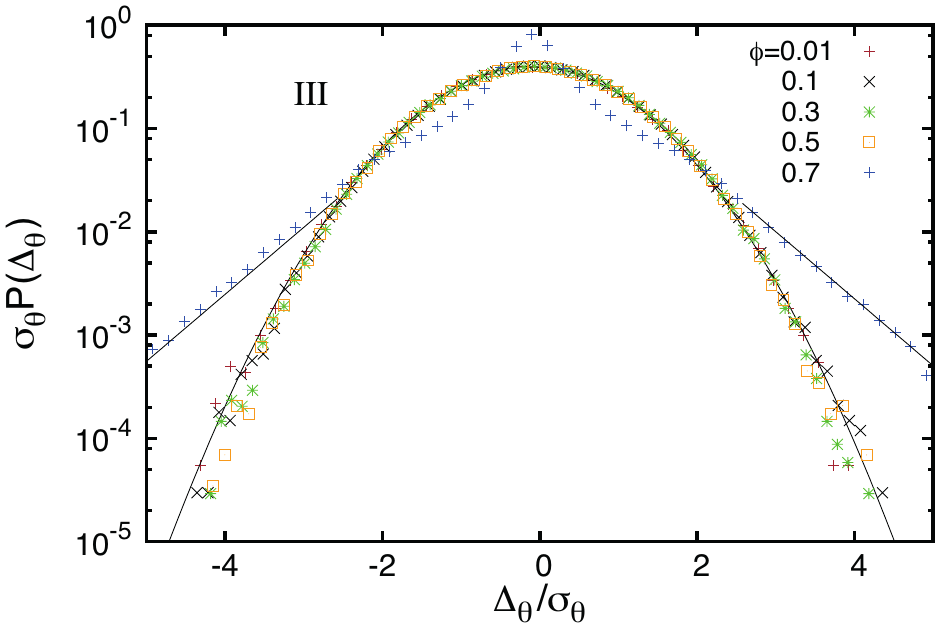}
  }
\caption{The pdf of the angular displacement 
at a fixed $\Delta t=t_a/2$ for which the total centre of mass displacement is non-monotonic, 
see Fig.~\ref{fig:MSD-CM-phase-coexistence}. Pe = 40 and densities 
given in the key. $\sigma_\theta$ is the dispersion of this pdf.
}
\label{fig:pdf-angle-phase-coexistence}
\end{figure}

The qualitative change in the global dynamics 
 is a consequence of the co-existence of dumbbells that move very differently 
in the sample, a fact that is confirmed by the analysis of the probability distributions of the individual dumbbell centre of mass and 
angular displacements displayed in Fig.~\ref{fig:pdf-CM-phase-coexistence}
and Fig.~\ref{fig:pdf-angle-phase-coexistence}. 
These figures, constructed for a fixed value of time 
delay $\Delta t=t_a/2$ (the III regime in the single molecule limit), present in fact
wide tails at high density $\phi=0.7$, strongly deviating from the Gaussian behaviour.

%Similar effects are found at smaller values of Pe, when getting close and within the 
%region of the phase diagram with phase co-existence. However, since the density
%interval in which there is phase co-existence is so narrow, it is better to 
%illustrate the phenomenon choosing a higher Pe value. 

\begin{figure*}[h!]
\begin{center}
  \begin{tabular}{ccccc}
	\resizebox{0.65\columnwidth}{!}{ \includegraphics{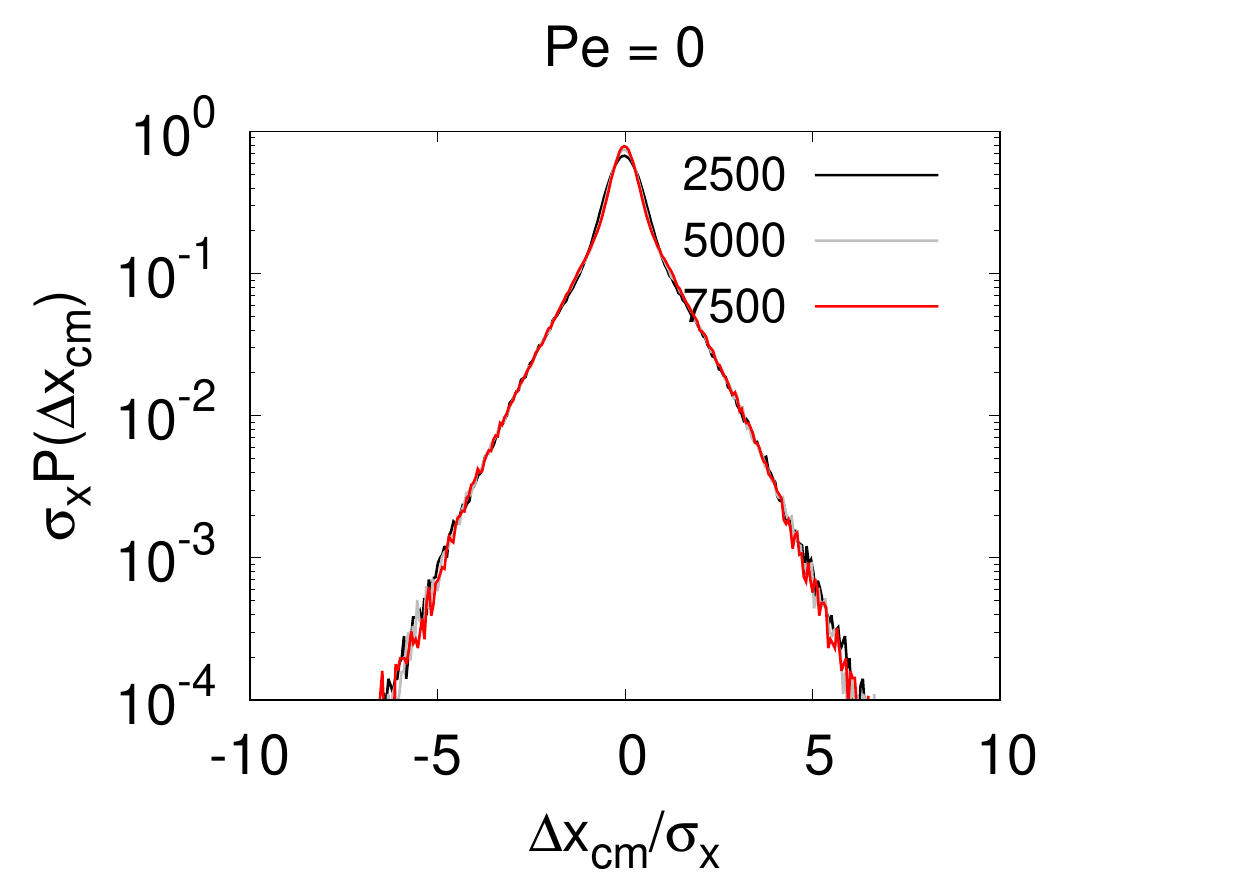}}
	\resizebox{0.65\columnwidth}{!}{ \includegraphics{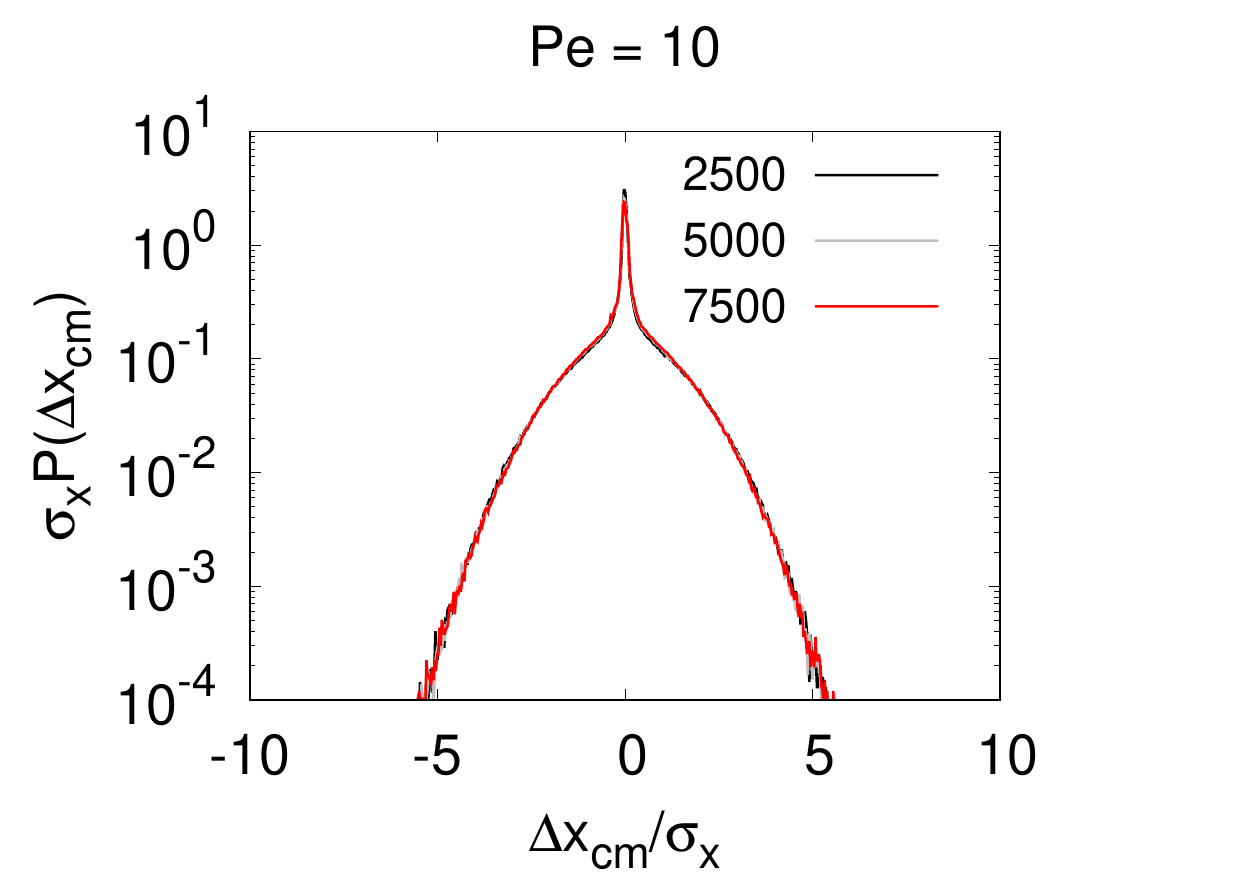}}
	\resizebox{0.65\columnwidth}{!}{ \includegraphics{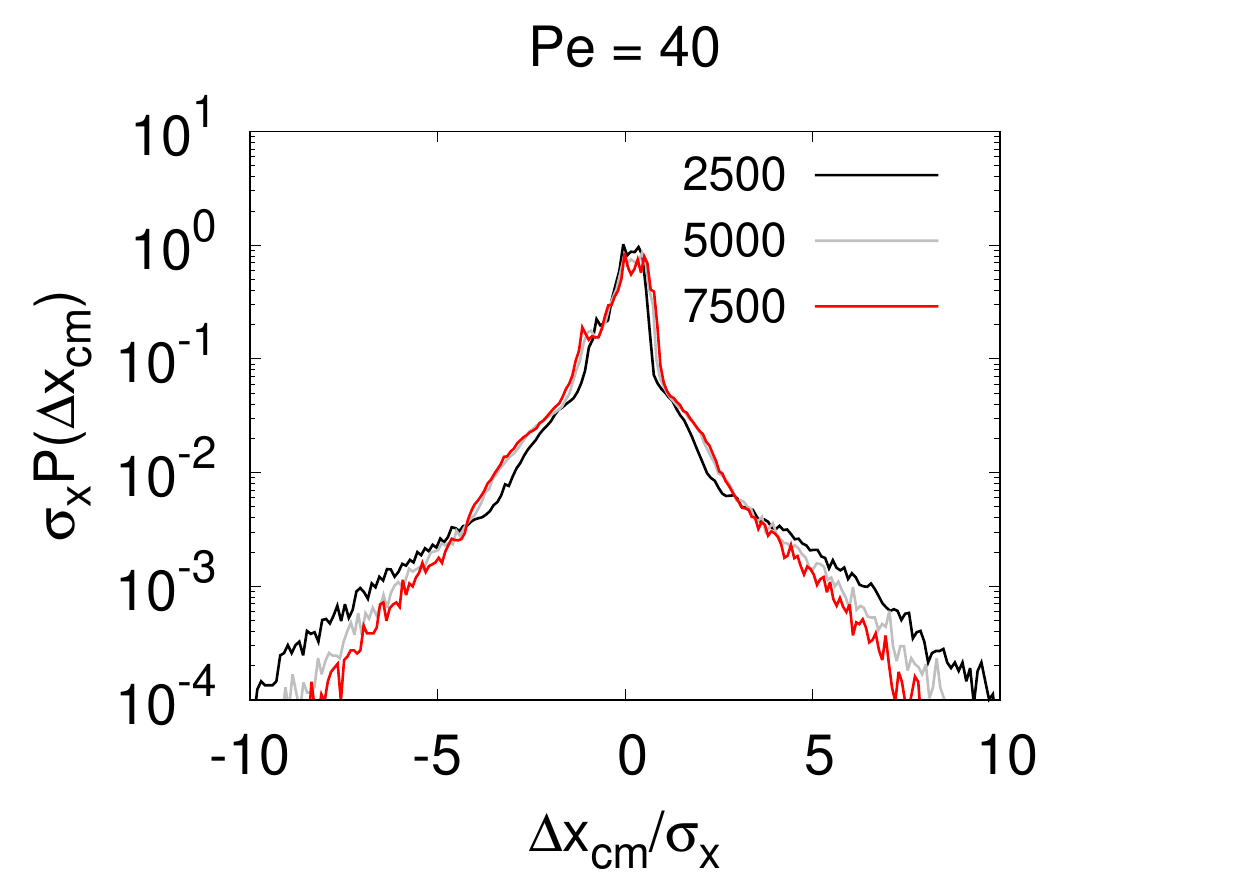}}
    \end{tabular}
  \caption{Distribution of individual centre of mass displacements 
at Pe = $0$, $10$, $40$ in the $50\%-50\%$ line at 
$\Delta t = 2500, \, 5000, \, 7500$, with $\Delta t>t_a=100$ in the IV regime. Log-lin scale is used and the pdfs are normalised using the mean-square-displacement of the raw histograms.
}
\label{fig:displ50-50}
\end{center}
\end{figure*}

We can check if the presence of wide tails is a general feature across different Pe inside the co-existence region, by following the distributions of individual centre of mass displacements on the 50\%-50\% curve in the phase diagram, see Fig.~\ref{fig:displ50-50}. Indeed, one finds a consistent behaviour at all Pe and even at the longest 
time delays considered, $\Delta t>t_a=100$ (in the IV regime). This is also true on the  25\%-75\% line (not shown).

\begin{figure*}[h!]
\begin{center}
  \begin{tabular}{ccccc}
	\resizebox{0.65\columnwidth}{!}{ \includegraphics{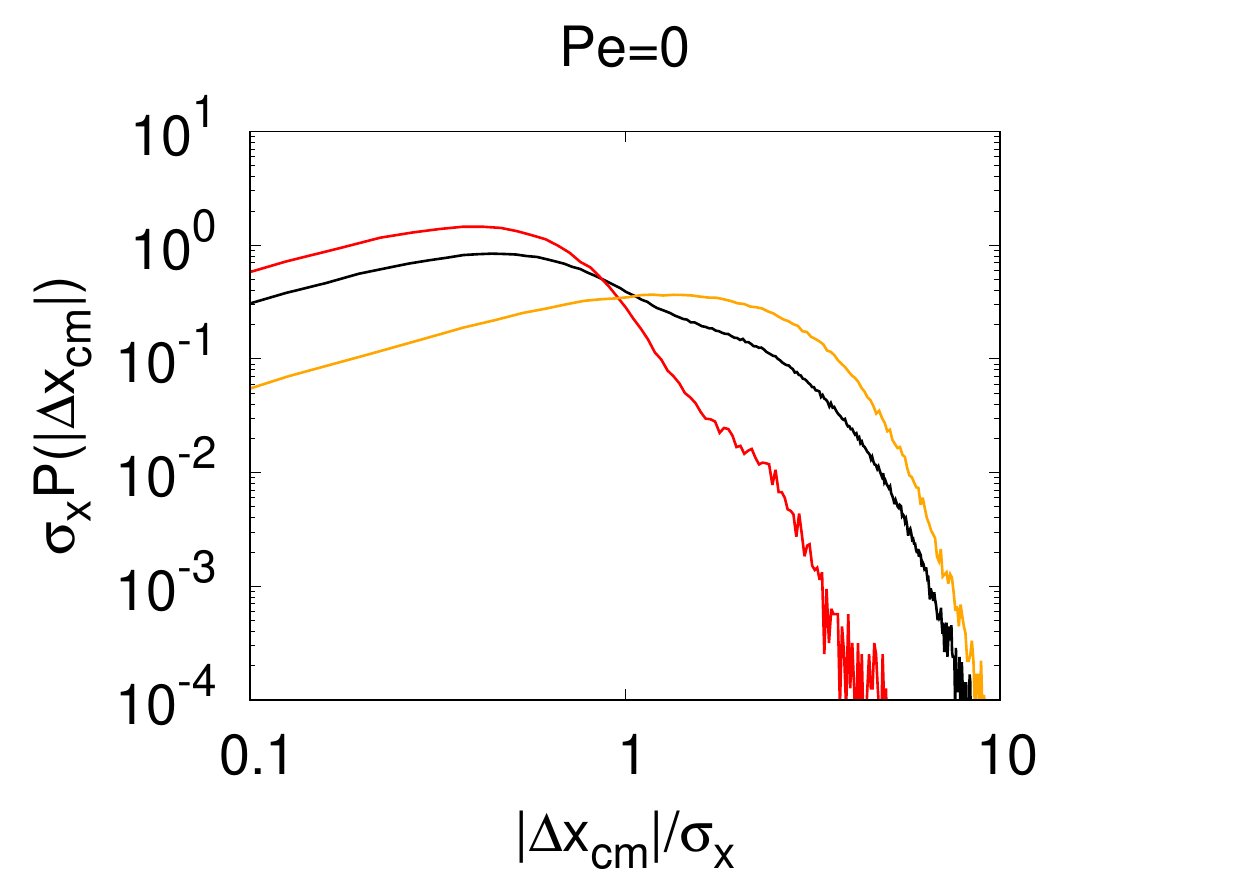}}
	\resizebox{0.65\columnwidth}{!}{ \includegraphics{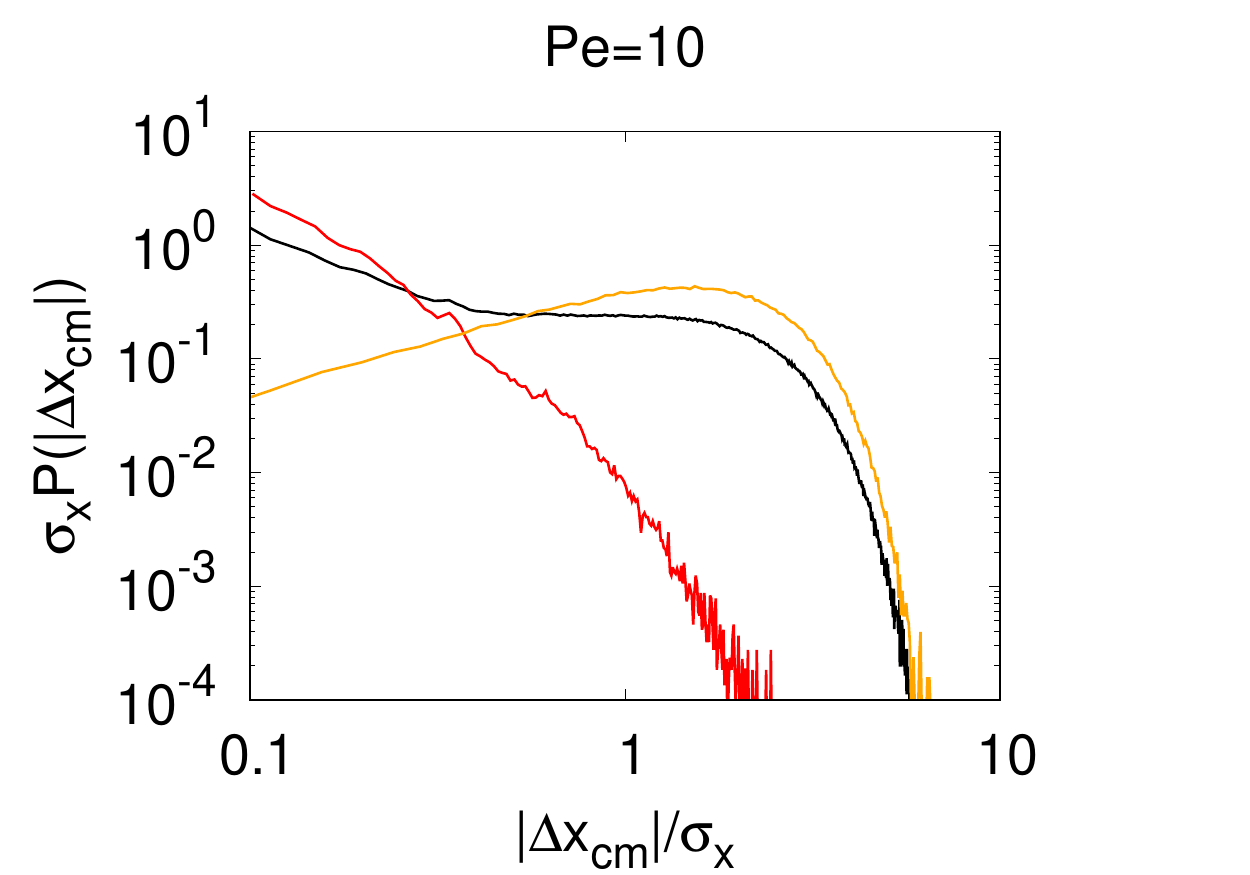}}
	\resizebox{0.65\columnwidth}{!}{ \includegraphics{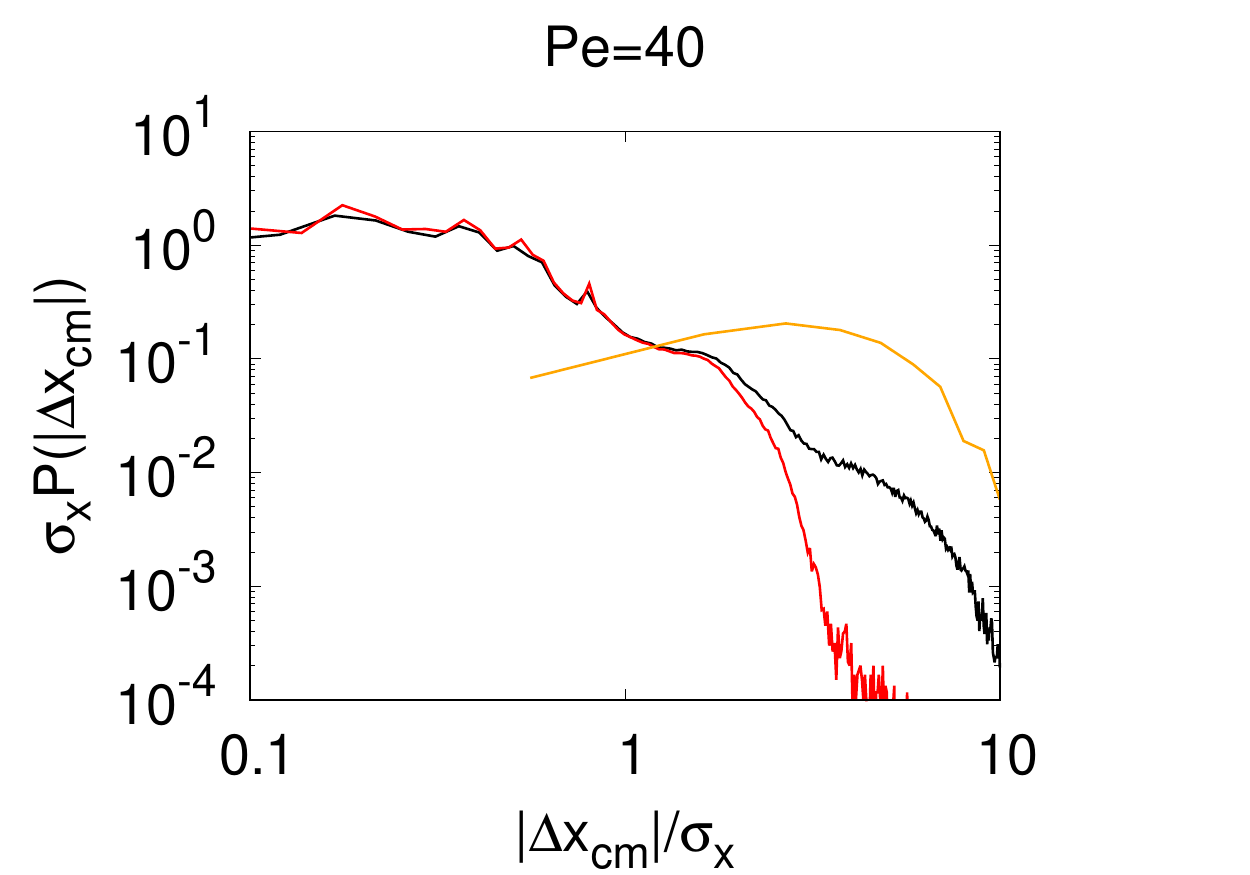}}
    \end{tabular}
  \caption{Separation of displacement modulus (black) into the cluster contribution (red) and liquid part (orange), 
at Pe = $0$, $10$, $40$  
in the $50\%-50\%$ at $\Delta t = 2500$. Log-log scale is used and the pdfs are normalised using the mean-square-displacement of the raw histograms.
}
\label{fig:displ-sep}
\end{center}
\end{figure*}

In order to interpret the wide tails we can measure separately the dumbbells displacements inside and outside clusters. The results of these  measurements are shown for 
the 50\%-50\% line in Fig.~\ref{fig:displ-sep}, where  distributions of the modulus of the displacements are  plotted.
With continuous black lines we plot the data for all beads, with dashed orange lines the data for dumbbells that are in aggregated
regions during the whole interval $\Delta t$ and, finally, with dotted red lines the data for dumbbells that are in liquid regions during
the whole interval $\Delta t$ (dumbbells that change character in between the two measuring times are excluded from the statistics). We established that a particle belongs to a cluster whenever the averaged hexatic modulus on a disk of radius
$R=$ 10 $\sigma_{\rm d}$ centred around the particle itself is greater than 0.75. 
%Only such particles were taken into account in the red peak on the right.
\begin{itemize}
\item[$\bullet$]
%The distributions of the modulus of the displacements are shown in Fig.~\ref{fig:displ-sep} on the on the 50\%-50\% curve in the phase diagram. 
The global displacement distributions  are  characterised by two separate peaks at all Pe, that appear to be directly linked to the separate distributions of displacements for particle inside and outside clusters, which are instead single-peaked. The peak value for each separate distribution increases monotonically with Pe, confirming the absence of any discontinuity in the region of phase coexistence  for all Pe~\cite{Cugliandolo2017}.
\end{itemize}

\subsection{The effective temperature}

\begin{figure}
\resizebox{\columnwidth}{!}{
  \includegraphics{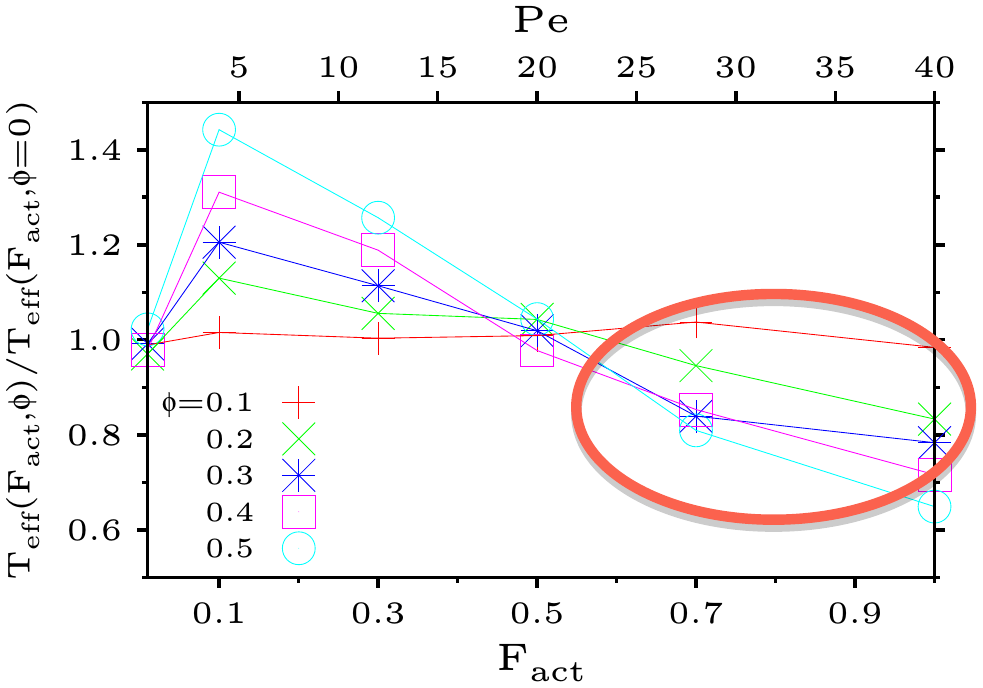}
  }
\caption{The global effective temperature as a function of the Pe number (scale above) or the 
active force (scale below) in a system at $T=0.05$ with various global densities given in the key. See the text for a discussion of the 
change in trend at around Pe = 5.
}
\label{fig:Teff}
\end{figure}

Figure~\ref{fig:Teff} shows the global effective temperature defined in Eq.~(\ref{eq_teff}) as a
function of the Pe number (scale above) or the 
active force (scale below) in a system at $T=0.05$ with around $N=10^5$ dumbbells~\cite{Suma14b}, leading to the various global densities given in the key. 
The combined dependence on density and Pe changes for Pe values beyond Pe = 5, approximately, suggesting that a finer analysis of the 
behaviour of the individual dumbbells is also needed in this case. Indeed, the (very) naive expectation that higher Pe means more agitation and 
hence a higher effective temperature coincides with the measurements of $T_{\rm eff}$ for Pe $\leq 5$, approximately.
\begin{itemize}
\item[$\bullet$]
 Beyond  Pe $\simeq 5$ the 
trend reverses and $T_{\rm eff}$ diminishes with increasing Pe for $\phi > 0.2 $, say. These are roughly parameters where one can expect 
some aggregates start appearing. No such inversion was found in the study of the FDR $T_{\rm eff}$ in the {\it homogeneous} atomic and 
molecular active liquids studied in~\cite{cugl-mossa1,cugl-mossa2,cugl-mossa3} nor in the model used in~\cite{LevisBerthier15}.
\end{itemize}

Indeed, one can expect strong differences in the value of the effective temperature
measured for dumbbells that are and remain within a cluster in the period $\Delta t$ over which the measurement is performed, 
and those that are free to move as a fluid during the same period. Of course, there are dumbbells that are free at the first measuring time 
and get trapped at the later one or, 
vice versa, are trapped and then let free during the internal $\Delta t$. A careful analysis should be carried out to understand the 
effect of these strong heterogeneities (see, {\it e.g.} the discussion in~\cite{ChCu07} for study of fluctuations in 
glassy systems and how these may affect the effective temperature defined from the fluctuation dissipation relation).

Certain confusion reigns in the active matter field concerning the identification of an effective temperature.

We first recall that the $T_{\rm eff}$ defined from the FDR satisfies many thermodynamic properties in glassy systems, 
at least for simple models that have been studied both analytically and numerically, and it is therefore a good candidate 
to play a similar role in other non-equilibrium cases. 

The effective temperature derived form the fluctuation-dissipation relation (FDR) {\it does not} represent an instantaneous configuration 
of the system. Therefore, it does not make sense to extract $T_{\rm eff}$ from the comparison between a 
non-equilibrium configuration and an equilibrium one at $T_{\rm eff}$ as done in~\cite{Fily12}. This fact is well-known in the 
field of glasses, where it was understood that the relaxation of the system {\it at different time-scales} has to be taken seriously in the measurement and 
interpretation of the effective temperature. (Trying a comparison between a glassy and an equilibrium configuration would also lead to meaningless statements
in this field.) Instead, the effective temperature defined from the FDR is a dynamic concept, obtained from the comparison of configurations 
visited at (very) different times with and without perturbation. 

It is also important to stress that $T_{\rm eff}$ {\it is not} proportional to the 
variance of an effective thermal noise in a Langevin equation for a reduced system obtained after integrating away some variables, 
due to the fact that the interactions remaining in the reduced system also play a role in the value that  $T_{\rm eff}$ will take. 
Typically, the (mean-field) single-variable Langevin equations that one derives from such a reduction have memory, are non-linear and the 
effective noise correlation depends also on the correlation functions of the selected variables of the reduced system (their mode-coupling 
version can be found in~\cite{ck:00}, for example, together with a discussion of how one should proceed to identify the effective temperature
from a non-trivial stationary distribution  in some complex systems). No single parameter, to associate to a temperature, stems out  immediately 
from the stochastic equation ruling the dynamics of the reduced system. It should then be no surprise that 
the temperature extracted from the variance of the effective noise arising from such a 
reduction in an active Brownian model was pathological~\cite{Fily12}. The ensuing contradictions found attached to this definition 
were interpreted as evidence against the existence of an effective temperature for active matter. In our opinion, these contradictions are the simple 
consequence of having used an erroneous definition, one that is already known to fail in other out of equilibrium systems.

Under which conditions $T_{\rm eff}$ extracted from the FDR may have a thermodynamic interpretation
also in active matter is an important issue. 
A quite natural condition seems to be that the out of equilibrium dynamics should approach equilibrium either 
in a long-time limit or by taking a parameter that controls the detailed balance break-down to zero ({\it e.g.},
in sheared liquids, taking the shear to vanish). Models that do not have such an equilibrium limit, like the trap model 
for glasses or various unbounded diffusion problems, have $T_{\rm eff}$s extracted from the FDRs with numerous 
anomalies~\cite{Fielding02}. One can then infer that models of active matter with no equilibrium limit should also have trouble in having a 
{\it bona fide} effective temperature. The model we studied in this paper has a natural passive limit that consists in 
simply setting the active forces to zero. The model in which particles are propelled by persistent noises studied 
in, {\it e.g.}~\cite{fodor16}, also allows for equilibrium in the limit of vanishing persistence time. Consequently,  these two families 
of models are good candidates to have meaningful effective temperatures. This is not the case 
of all models in the literature.

In particular, it is important to evaluate whether the effective temperature found {\it via} the use of the (time - delayed) 
FDR for different observables evolving in the same regime take the same values. This question was answered positively, for some observables, 
in~\cite{cugl-mossa1,cugl-mossa2,cugl-mossa3}
where very simple active models for particles and polymers were used. The results in~\cite{LevisBerthier15}, however, 
seem to point in a different direction in a model of self-propelled hard-disks. In this paper, a length-dependent
effective temperature in the dilute and liquid regimes was found, while these different values are locked to a single unique one 
in the glassy (high density) regime. Whether partial equilibrations between different degrees of freedom 
occur in active systems might then depend on the model and/or on the parameters and associated dynamic regime.

Another property of the effective temperature defined from the FDR in glassy systems is that it is related
to the configurational entropy or complexity, as found in mean-field solvable models and also in a number of 
numerical simulations. A very interesting investigation along these lines in the context of active matter
appeared recently in~\cite{PreislerDijkstra16}.

The effective temperature $T_{\rm eff}$ should be measurable with a thermometer. The idea of using (tuned) tracer particles as 
thermometers is described in~\cite{cugl:review} where references to papers where these tests have been 
performed in glassy systems can be found. In the context of active matter, following~\cite{BenIsaac15}, a very recent preprint explores
this issue~\cite{Gov18}:  the kinetic energy of a passive tracer particle is shown to coincide with the
$T_{\rm eff}$ measured in the zero frequency (or long time delay) limit.

Finally, we simply want to mention that several papers in the active matter literature use the 
FDR to estimate the effective temperature of the system both analytically~\cite{Tailleur09,Wang11,Szamel14} and 
experimentally~\cite{Martin01,Mizuno07,Ben-Isaac-etal,Palacci10}.

\section{Discussion}
\label{sec:open-pbms-dyn}

In this work we revisited the structural and dynamical properties of an ensemble of 
active dumbbells in repulsive interaction through a quite hard potential. With the study of new 
observables, we confirmed the
coexistence between liquid and hexatic order in a narrow interval of packing 
fraction densities of the passive system, that progressively extends into the Pe $> 0$ part of the 
phase diagram, with no discontinuity.

We quantified the separation in dilute and dense regions, the latter with hexatic order, 
by looking at the probability distribution functions of local densities and local hexatic order
and correlating the two. The distribution of the centre of mass MSD also showed the existence of 
two distinct populations of dumbbells.

As new results, we found that the mobility of the dumbbells confined to clusters is highly suppressed.
For strong enough Pe, say Pe ${\geq 50}$, we showed that the spontaneously formed clusters turn around their centre of mass 
with an angular velocity that is proportional to the inverse of their radii. The poly-crystalline nature
of the clusters, with respect to the hexatic order, does not seem to play a major role in their rotational 
properties. Instead, the orientation of the dumbbells inside the clusters is, indeed, important, as a certain amount of 
disorder is needed to make them turn. The enstrophy pdfs allowed us to quantify the rotating properties of the 
individual clusters.

The possible influence of the systems' heterogeneity on the effective temperature measurements was also 
stressed in the text. A more detailed analysis, separating contributions from dumbbells confined to 
clusters, to freely displacing ones in the dilute phase, would be needed to justify whether the decrease of 
$T_{\rm eff}$ with increasing Pe, in dense systems, is due to the macroscopic dumbbell aggregation.

We stress the fact that we used a rather hard repulsive potential with power law decay 
characterised by a short-distance power law decay with $2n=64$. Recent studies by Klamser {\it et al.}~\cite{KlKaKr18} 
on active disks interacting {\it via} a repulsive potential with $n=3$, for which the passive limit complies with 
the standard HNY melting scenario, point towards a different phase diagram. We shall come back to 
the behaviour of active Brownian disks interacting {\it via} a very hard repulsive potential in a separate publication~\cite{DigrLeCuGoPaSu18}.

Here and in~\cite{Cugliandolo2017} we analysed the structure and dynamics of the active molecular 
system. Unfortunately, we cannot utilise an equation of state for active dumbbells since no unambiguous 
definition of pressure exists for these systems. Finding such a definition remains a very interesting open 
problem.

All our studies concern mono-disperse systems where structural arrest of glassy type is not expected. Recent work 
by Mandal {\it et al.} focuses on the effects of poly-dis\-pers\-ity and how super-cooled liquid facts 
and glassiness~\cite{Chong2005,Moreno2005,Chong2009} manifest in active dumbbell systems. We refer the reader 
to this reference for further details~\cite{Mandal2017a,Mandal2018}.

\vspace{0.5cm}

\noindent
{\bf Acknowledgements.} 
Simulations ran on IBM Nextscale GALILEO at CINECA (Project INF16-fieldturb) under CINECA-INFN agreement and 
at Bari ReCaS e-Infra\-struc\-ture funded by MIUR through PON Research and Competitiveness 2007-2013 Call 254 Action I.
GG acknowledges MIUR for funding (PRIN 2012NNRKAF). LFC is a member of Institut Universitaire de 
France, and thanks the KITP University of Santa Barbara for hospitality during part of the preparation of this work and 
L. Berthier, P. Choudhuri, C. Dasgupta, M. Dijkstra, J. Klamser, D. Levis , R. Mandal and J. Palacci for very useful discussions. 

\noindent
All authors conceived the study. P.D. and I.P. performed the simulations and produced the figures. All authors commented and wrote the paper.
 
% BibTeX users please use
\bibliographystyle{siam-order-appearance}
\bibliography{dumbbells-biblio}

\end{document}